\documentclass[twocolumn]{aastex63}

\newcommand{\revise}[1]{{#1}}
\newcommand{\editone}[1]{{#1}}
\received{\today}
\revised{}
\accepted{}

\shorttitle{AGN as DHOs}
\shortauthors{Yu et al.}

\newcommand{\sigmadho}{$\sigma_{\mathrm{DHO}}$}
\newcommand{\sigmadrw}{$\sigma_{\mathrm{DRW}}$}
\newcommand{\sigmanoise}{$\sigma_{\mathrm{\epsilon}}$}
\newcommand{\tauperturb}{$\tau_{\mathrm{perturb}}$}
\newcommand{\taudecay}{$\tau_{\mathrm{decay}}$}
\newcommand{\taudecorr}{$\tau_{\mathrm{decorr}}$}
\newcommand{\taurise}{$\tau_{\mathrm{rise}}$}

\newcommand{\lledd}{$L/L_{\mathrm{edd}}$}
\newcommand{\lbol}{$L_{\mathrm{bol}}$}
\newcommand{\mbh}{$M_{\mathrm{BH}}$}
\newcommand{\lambdarf}{$\lambda_{\mathrm{RF}}$}
\newcommand{\FeII}{$\mathrm{Fe\,II}$}
\newcommand{\CIV}{$\mathrm{C\,IV}$}

\newcommand{\OIII}{$\mathrm{[O\,III]}$}
\newcommand{\Rg}{$R_{\mathrm{g}}$}

\newcommand{\rperturb}{$r_{\mathrm{perturb}}$}
\newcommand{\maxdt}{$\mathrm{Max}\,\Delta t$}
\newcommand{\mindt}{$\mathrm{Min}\,\Delta t$}

\begin{document}

\title{Examining AGN UV/optical Variability Beyond the Simple Damped Random Walk}

\correspondingauthor{Weixiang Yu}
\email{wy73@drexel.edu}

\author[0000-0003-1262-2897]{Weixiang Yu}
\affiliation{Department of Physics, \\
Drexel University, 32 S.\ 32nd Street, \\
Philadelphia, PA 19104, USA}

\author[0000-0002-1061-1804]{Gordon T. Richards}
\affiliation{Department of Physics, \\
Drexel University, 32 S.\ 32nd Street, \\
Philadelphia, PA 19104, USA}

\author[0000-0001-7416-9800]{Michael S. Vogeley}
\affiliation{Department of Physics, \\
Drexel University, 32 S.\ 32nd Street, \\
Philadelphia, PA 19104, USA}

\author[0000-0001-9134-6522]{Jackeline Moreno}
\affiliation{Department of Physics, \\
Drexel University, 32 S.\ 32nd Street, \\
Philadelphia, PA 19104, USA}

\author[0000-0002-3168-0139]{Matthew J. Graham}
\affiliation{Department of Physics, Math, and Astronomy\\
California Institute of Technology\\
Pasadena, CA, 91125, USA}

\begin{abstract}
We present damped harmonic oscillator (DHO) light-curve modeling for a sample of 12,714 spectroscopically confirmed quasars in the Sloan Digital Sky Survey Stripe 82 region. DHO is a second-order continuous-time autoregressive moving-average (CARMA) process, \revise{which can be fully described using} four independent parameters: a natural oscillation frequency ($\omega_{0}$), a damping ratio ($\xi$), a characteristic perturbation timescale ($\tau_{\mathrm{perturb}}$), and an amplitude for the perturbing white noise ($\sigma_{\mathrm{\epsilon}}$). The asymptotic variability amplitude of a DHO process is quantified by $\sigma_{\mathrm{DHO}}$---a function of $\omega_{0}$, $\xi$, $\tau_{\mathrm{perturb}}$, and $\sigma_{\mathrm{\epsilon}}$.
We find that both $\tau_{\mathrm{perturb}}$ and $\sigma_{\mathrm{\epsilon}}$ follow different dependencies with rest-frame wavelength ($\lambda_{\mathrm{RF}}$) on either side of 2500 \AA, whereas $\sigma_{\mathrm{DHO}}$ follows a single power-law relation with $\lambda_{\mathrm{RF}}$.
After correcting for wavelength dependence, $\sigma_{\mathrm{DHO}}$ exhibits anti-correlations with both the Eddington ratio and the black hole mass, \revise{while} $\tau_{\mathrm{perturb}}$---with a typical value of days in the rest-frame---shows an anti-correlation with the bolometric luminosity. 
Modeling AGN variability as a DHO offers more insight into \revise{the workings of accretion disks} close to the supermassive black holes (SMBHs) at the center of AGN. The newly discovered short-term variability (\revise{characterized by $\tau_{\mathrm{perturb}}$ and $\sigma_{\mathrm{\epsilon}}$}) and its correlation with bolometric luminosity pave the way for new algorithms that will derive fundamental properties (e.g., Eddington ratio) of AGN using photometric data alone. 

\end{abstract}


\section{Background \& Motivation} \label{sec:intro}
The UV/optical luminosity of active galactic nuclei (AGN)\footnote{In this manuscript, we will use AGN and quasar interchangeably without making a distinction based on, e.g., luminosity or radio emission.}, or quasars at the bright end, is known to vary at the 10\% flux level on average from weeks to years \citep{vandenberk2004, sesar2007}.
The time variability of AGN luminosity has been known for decades~\citep{Matthews1963}, however, the physical mechanisms driving such variability are still unclear. Nevertheless, the success of reverberation mapping~\citep{peterson2004} has shown that the broad emission lines respond to and lag behind the continuum fluctuations, suggesting an accretion disk origin of the UV/optical continuum variability.

Under the assumption that the optical variability originates from the accretion disk close to the SMBH, various models have been proposed to explain the observed variability. Based on the standard $\alpha$ disk model~\citep{shakura1973}, some have shown that the observed optical variability could be driven by variations in the mass accretion rate~\citep[][]{Pereyra2006, li2008, liu2008}. Meanwhile, others suggested that it is also possible for the accretion disk to passively reprocess the radiation from the X-ray corona given the observed short time lags between UV/optical continuum light curves~\citep[][]{wanders1997, sergeev2005}.

Observationally, significant efforts/progress have also been made to investigate the physical origin(s) of the UV/optical variability by virtue of exploring the correlations of variability signatures with the physical properties of AGN.
Notable results include: an anti-correlation of variability amplitude with rest\revise{-frame} wavelength and an anti-correlation of variability with luminosity and/or Eddington ratio~\citep[e.g.,][]{vandenberk2004, wilhite2007, bauer2009, macleod2010, simm2016, caplar2017}. Correlations between variability and black hole mass have also been reported in numerous studies~\citep[][]{wold2007, wilhite2007, bauer2009, kelly2009, macleod2010, caplar2017}. 

\revise{Among the different techniques utilized to characterize AGN variability}, \citet{kelly2009} started a new era of directly modeling (inherently non-periodic) AGN light curves using stochastic diffusion processes, in particular, as a damped random walk (DRW) model. The DRW model features a fixed-slope power spectrum density (PSD) at high frequencies (short timescales) and a flat PSD at timescales longer than a characteristic timescale ($\tau_{\mathrm{DRW}}$).  
Modeling AGN variability as a DRW has provided a lot of insight into how AGN luminosity varies in UV/optical and what might be driving it \citep{macleod2010}. 
However, better sampled light curves from the Kepler Mission \citep{kepler2010} cast doubt on the DRW description of AGN variability because of the steeper slopes observed in the PSDs at high frequencies \citep{mushotzky2011}; investigations carried out by other groups also arrived at similar conclusions \citep{kasliwal2015, simm2016, smith2018}. 
This discrepancy motivates the search for new models (and methods) to analyze AGN light curves. 
Given that the DRW model is the simplest case of a more general class of stochastic diffusion processes, namely the continuous-time autoregressive moving-average processes \citep[CARMA;][]{brockwell2001, rouxalet2002}, \citet{kelly2014} set up a more flexible framework to model astronomical time series as CARMA processes, where the PSDs of higher-order CARMA processes can take more flexible shapes, for example, a wide range of PSD slopes can be achieved at high frequencies.
Later, \citet{kasliwal2017} demonstrated that the CARMA(2,1) model is a better fit than all other models of CARMA for a well-monitored object Zw229-15. 
Inspired by the aforementioned discrepancy and the pilot investigations carried out by \citet{kelly2014} and \citet{kasliwal2017}, \citet[][hereafter M19]{moreno2019} conducted an in-depth exploration of the CARMA(2,1) model, otherwise known as the (perturbation-driven) damped harmonic oscillator (DHO) model, and \revise{established} guidelines for modeling AGN light curves as DHOs. 

Here, we build upon the work performed by \citet{kelly2014}, \citet{kasliwal2017}, and \citet{moreno2019} to model a large statistical sample of AGN as DHOs, examine the variability signatures extracted by the DHO model, and explore the potential correlations between DHO parameters and the physical properties of AGN. We acknowledge that the DHO model (alike the DRW model) is a statistical model rather than a physical model; however, stochastic diffusion processes such as CARMA are a natural choice for parameterizing AGN light curves,
and it can reveal interesting variability features embedded in the light curves that could not be unfolded otherwise~(see \citet{vio2018} for a discussion about the limitations of CARMA modeling). 

In Section~\ref{sec:data}, we introduce the data set utilized in this work and outline our initial light curve construction procedures. In Section~\ref{sec:method}, we provide an overview of the DHO model, its key features, and how to use Gaussian process (GP) to \revise{extract DHO parameters from} light curves. In Section~\ref{sec:best_fit}, we present the results of fitting DHO to our quasar light curves, layout and test our bad-fit identification algorithm, and explore the effects of light-curve sampling and photometric accuracy on the best-fit DHO parameters. 
\revise{In Section~\ref{sec:results}, we describe the observed correlations of DHO features with the physical properties of the quasars in our sample, and discuss the associated implications in Section~\ref{sec:discussion}.}
Finally, we summarize our results and provide an outlook for future work in direct modeling of AGN light curves using stochastic diffusion processes in Section~\ref{sec:Summary}.

\section{The Data Set} \label{sec:data}
\begin{figure}
    \centering
    \includegraphics[scale=0.4]{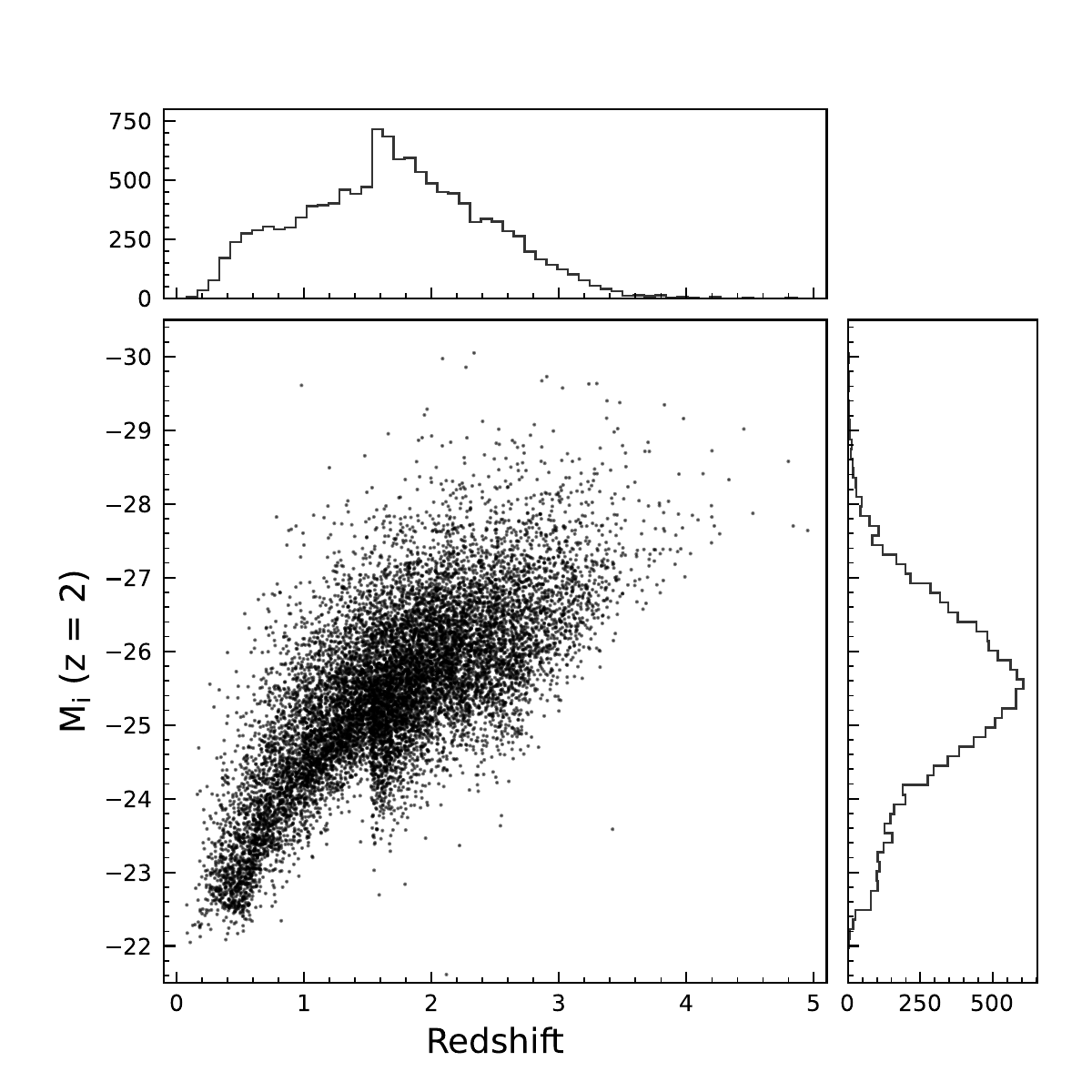}
    \caption{Distribution of the \textbf{main sample} quasars in the luminosity (absolute $i$-band magnitude k-corrected to a redshift of 2) and redshift space.  This sample is constructed from the SDSS DR16 quasar catalog---the latest release of SDSS quasar spectra~\citep{lyke2020}; quasars in this sample either have their physical properties \editone{estimated} by \citet{shen2011} or \CIV\ emission line properties measured by \citet{rankine2020}.
    }
    \label{fig:m_z_dist}
\end{figure}

We compiled a sample of 12,714 spectroscopically confirmed quasars discovered in the Sloan Digital Sky Survey (SDSS) Stripe 82 region \citep{york2000a, annis2014}, a $120^{\circ}$ long and $2.5^{\circ}$ wide stripe centered along the celestial equator, from the quasar catalog of SDSS Data Release 16 \citep[DR16Q;][]{lyke2020}.
We refer to this initial sample of quasars as the \textbf{main sample}; quasars in this main sample either have their fundamental physical properties (e.g., black hole mass) \editone{estimated} by \citet{shen2011} or \CIV\ emission line properties (i.e., equivalent width and blueshift) \revise{determined} by \citet{rankine2020}. 
Figure~\ref{fig:m_z_dist} shows the distribution of the main sample quasars in the luminosity and redshift space, where the luminosity ($i$-band absolute magnitude) has been k-corrected to $z = 2$ \citep{richards2006a}.

\subsection{The Sloan Digital Sky Survey (2000--2008)}\label{subsec:data_sdss}
The Sloan Digital Sky Survey obtained images for more than $10,000\,\mathrm{deg}^2$ of the northern hemisphere down to limiting magnitudes of 22.5, 23.2, 22.6, 21.9, 20.8 at the 50$\%$ completeness level in the $u, g, r, i, z$ bands, respectively. 
The SDSS Stripe 82 (S82) region was observed repeatedly over a 8-year long baseline providing up to 90 single-epoch observations \citep{frieman2008,sako2008,abazajian2009}. 
The SDSS light curves used in this investigation span two phases of the Sloan Digital Sky Survey, namely, the SDSS Legacy Survey \citep{york2000a} and the SDSS-II Supernova Survey \citep{frieman2008, sako2008}, where the latter was performed under less photometric conditions. 
A so-called ``ubercalibration" \citep{ivezic2007, padmanabhan2008, bramich2008}, which takes advantage of the overlap between adjacent imaging runs to arrive at a uniform internal calibration, was utilized to achieve a $\sim$1\% photometric accuracy in $griz$ and 2\% in $u$ for both photometric and less photometric observations. 
Such a calibration method has become the default since the seventh data release of SDSS (DR7) \citep{abazajian2009}, from which our light curves were constructed.

\begin{figure*}[ht!]
    \centering
    \includegraphics[width=0.95\textwidth]{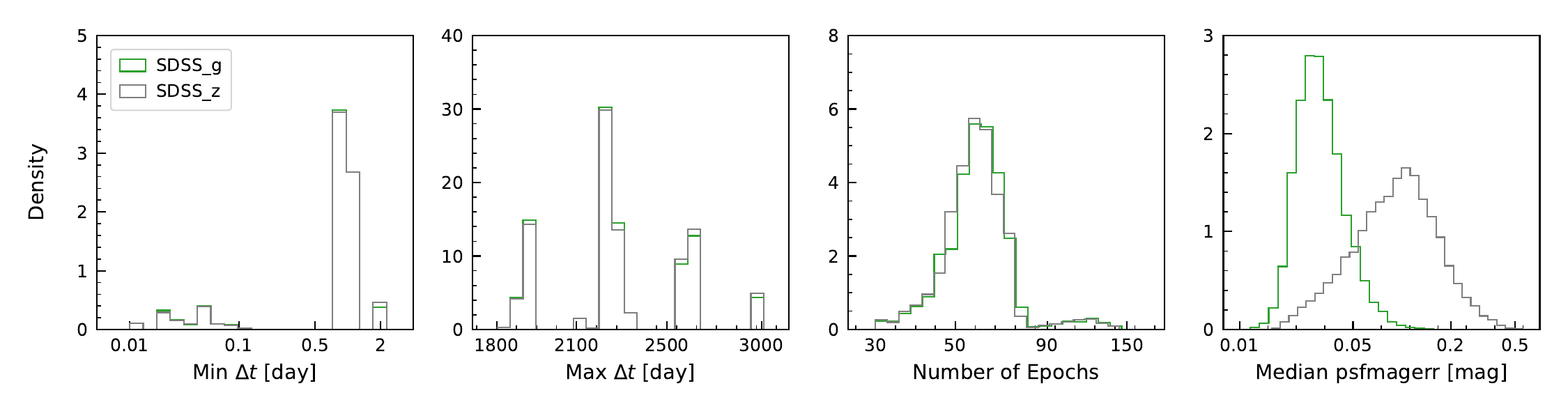}
    \vspace{-0.2cm}
    \caption{Distribution of SDSS light curve statistics in $g$ and $z$ bands. The statistics shown are (from left to right): minimum time separation between any two observations in a light curve (\mindt); the maximum time separation between any two observations in a light curve (\maxdt); number of epochs in a light curve; the median photometric error for observations in a light curve.}
    \label{fig:lc_stat_dist}
\end{figure*}

\subsection{SDSS Light Curves}\label{subsec:data_lc_merge}
The SDSS light curves for our quasars were generated by cross matching against the \texttt{photoobj} table under the \texttt{Stripe82} context on CasJobs\footnote{https://skyserver.sdss.org/casjobs/} using a 1" matching radius. We imposed an initial quality cut on the matched detections---that is, their photometry must be ``clean" and their photometric uncertainties (\texttt{psfmagerr}) must be smaller than 1 mag. More on the ``clean" flag (and other photometry flags) can be found on SDSS-IV's  website\footnote{https://www.sdss.org/dr16/tutorials/flags/}. The raw light curves are then post-processed following the recipe described below:
\begin{enumerate}
    \item We first require light curves to have at least 30 epochs.
    \item We then remove data points that deviate more than 3$\sigma$ away from the 3-point running median. This process removes any abnormally large-amplitude variability in the photometry~\citep{graham2014}. 
    \item Lastly, we remove data points with photometric uncertainties that are more than 5 times larger than the median uncertainty of all photometry in the corresponding light curves.
\end{enumerate}

Note that the post-processing described above is on a per-band basis, that is, failing to have a good light curve in a particular passband does not exclude an object from our sample. The final collection contains $\approx$12,400 light curves in each $u, g, r, i, z$ band. Those light curves are then fitted with the DHO model. The distributions of four basic statistics of those light curves are shown in Figure~\ref{fig:lc_stat_dist}.

\section{Methodology} \label{sec:method}

We model the time variability of AGN UV/optical luminosity as a DHO, which is \revise{formally} defined as the solution to the following stochastic differential equation,
\begin{equation}\label{eqn:dho1}
    d^{2} x+\alpha_{1} d^{1} x+\alpha_{2} x=\beta_{0} \epsilon(t)+\beta_{1} d^{1}(\epsilon(t)),
\end{equation}
where $\epsilon(t)$ is Gaussian white noise with an amplitude of unity\footnote{$\epsilon(t)\equiv dW/dt$, $W$ is the Wiener process or referred to as Brownian motion in physics.}, $\alpha_{1}$ and $\alpha_{2}$ are called the autoregressive (AR) coefficients\footnote{Note that $\alpha_0$ is defined to be 1 by convention.}, and $\beta_{0}$ and $\beta_{1}$ are called the moving-average (MA) coefficients. \revise{The differentiation is with respect to time.} Here, $x$ represents the brightness (magnitude\footnote{\editone{The output of a CARMA process follows a Gaussian distribution, so does the magnitude light curve of a compact accreting object~\citep{uttley2001, gaskell2004, macleod2012}, thus fitting CARMA models to quasar light curves in magnitude rather than in flux is a logical choice. Meanwhile, for the majority of our sample (dominated by luminous quasars), host galaxy (and other background) star light contamination is negligible~\citep{shen2011}}.} in this work) of the modeled quasars. For a comparison, the stochastic differential equation for the DRW model---a CARMA(1,0) process---has the form\footnote{Note that $\beta_{0}$ is equivalent to $\sigma$ in \citet{kelly2009, kelly2014} and $\hat{\sigma}$ in \citet{macleod2010}.},  
\begin{equation}\label{eqn:drw}
    d^{1} x+\alpha_{1} x(t)=\beta_{0} \epsilon(t).
\end{equation}

\subsection{Damped Harmonic Oscillator (DHO)}\label{subsec:dho}
The utility of modeling quasar light curves as DHOs has been explored and discussed extensively in M19; here we provide a brief introduction to the DHO model under the framework of an impulse-response dynamical system.

In short, we can think of time series (or light curves in this context) following the stochastic differential equation (SDE) shown in Equation~\ref{eqn:dho1} as impulse-response dynamical systems. The left-hand-side (LHS) of the SDE describes how the systems respond to impulse perturbations (in the differential form) and the right-hand-side (RHS) describes how the systems are being perturbed/excited. This interpretation connects the DHO process naturally to the classical damped harmonic oscillator, where the classical counterpart has a deterministic driving force on the RHS rather than a stochastic one. Given this analogy, we can rewrite Equation~\ref{eqn:dho1} as,
\begin{equation}\label{eqn:dho2}
    d^{2} x+2 \xi \omega_{0} d^{1} x+\omega_{0}^{2} x= \sigma_{\mathrm{\epsilon}} \epsilon(t)+\tau_{\mathrm{perturb}}\,\sigma_{\mathrm{\epsilon}}d^{1}(\epsilon(t)),
\end{equation}
where $2\xi\omega_{0} = \alpha_{1}$ and $\omega_{0}^{2} = \alpha_{2}$. $\xi$ is the damping ratio of the damped oscillator and $\omega_{0}$ is the natural oscillation frequency (i.e., when there is no damping). 
We can classify DHO processes into underdamped ($\xi < 1$) and overdamped ($\xi > 1$) DHOs, each corresponds to a different class of dynamical systems that can be interpreted using the impulse-response framework \citep{moreno2019}.
On the RHS of Equation~\ref{eqn:dho2} (compared to Equation~\ref{eqn:dho1}), we renamed $\beta_0$ to \sigmanoise\ and defined \tauperturb\ as $\beta_1/\beta_0$. \revise{\sigmanoise\ can be treated as the amplitude of the short-term perturbing white noise ($\epsilon(t)$) with the unit}\footnote{\revise{The unit of \sigmanoise\ can be derived by matching the units of the LHS and the RHS of Equation~\ref{eqn:dho2}~\citep{kasliwal2017} given that $\epsilon(t)$ ($\equiv dW/dt$) has the unit of $1/\sqrt{dt}$~\citep{rouxalet2002}.}} \revise{of $\mathrm{magnitude/time^{3/2}}$. 
For unit consistency, \tauperturb\ obtains a unit of time and manifests as a characteristic timescale of the perturbation process}~\citep{kasliwal2017, moreno2019}.

Given Equation~\ref{eqn:dho2}, we can fully define a DHO process using $\xi$, $\omega_0$, \sigmanoise, and \tauperturb, where the first two are from the LHS of Equation~\ref{eqn:dho2} and the last two are from the RHS of Equation~\ref{eqn:dho2}.
This new set of independent parameters and the four ($\alpha_1$, $\alpha_2$, $\beta_0$, and $\beta_1$) from Equation~\ref{eqn:dho1} will be used side by side throughout this manuscript, where the original parameters (from Equation~\ref{eqn:dho1}) will be used mostly in the technical sections (e.g., Section~\ref{sec:method}) and the newly defined parameters will be used primarily in the discussion of scientific implications (e.g., Section~\ref{sec:results}).
 
Various intrinsic timescales can be extracted from the LHS of Equation~\ref{eqn:dho1} (or Equation~\ref{eqn:dho2}) based on the roots ($r_{1}, r_{2}$) of its characteristic equation, 
\begin{equation}\label{eqn: dho_char}
    r_{1}, r_{2} = -\frac{\alpha_1}{2} \pm \sqrt{\frac{\alpha_{1}^2}{4} - \alpha_{2}} = -\omega_{0}\xi \pm \omega_{0}\sqrt{\xi^2 - 1}.
\end{equation}

When $\xi < 1$ (underdamped DHOs) the two roots are complex conjugates, a decay timescale ($\tau_{\mathrm{decay}}$) and a damped oscillation period ($T_{\mathrm{dQPO}}$) can be obtained,
\begin{eqnarray}
    \tau_{\mathrm{decay}} &&= \frac{1}{|\mathrm{Re} (r_{1})|} = \frac{1}{\omega_{0}\,\xi}\nonumber,\\
    T_{\mathrm{dQPO}} &&= \frac{2\pi}{|\mathrm{Im} (r_{1})|} 
    = \frac{2\pi}{\omega_{0}\sqrt{1 - \xi^2}}. 
\end{eqnarray}
where $2\pi/\omega_0$ is the natural oscillation period ($T_{\mathrm{QPO}}$) associated with $\omega_0$. $T_{\mathrm{dQPO}}$ is effectively the oscillation period in the presence of resistance/damping. For no resistance/damping ($\xi \approx 0$), $T_{\mathrm{dQPO}}$ and $T_{\mathrm{QPO}}$ become equivalent. 

When $\xi > 1$ (overdamped DHOs), both roots are real leaving us a rising timescale ($\tau_{\mathrm{rise}}$) and a decay timescale ($\tau_{\mathrm{decay}}$),
\begin{eqnarray}
    \tau_{\mathrm{rise}} = |\frac{1}{\mathrm{min}(r_{1}, r_{2})|},
    \quad
    \tau_{\mathrm{decay}} = |\frac{1}{\mathrm{max}(r_{1}, r_{2})|}.
\end{eqnarray}

\begin{figure*}[ht!]
    \vspace{-1cm}
    \centering
    \includegraphics[width=1.02\textwidth]{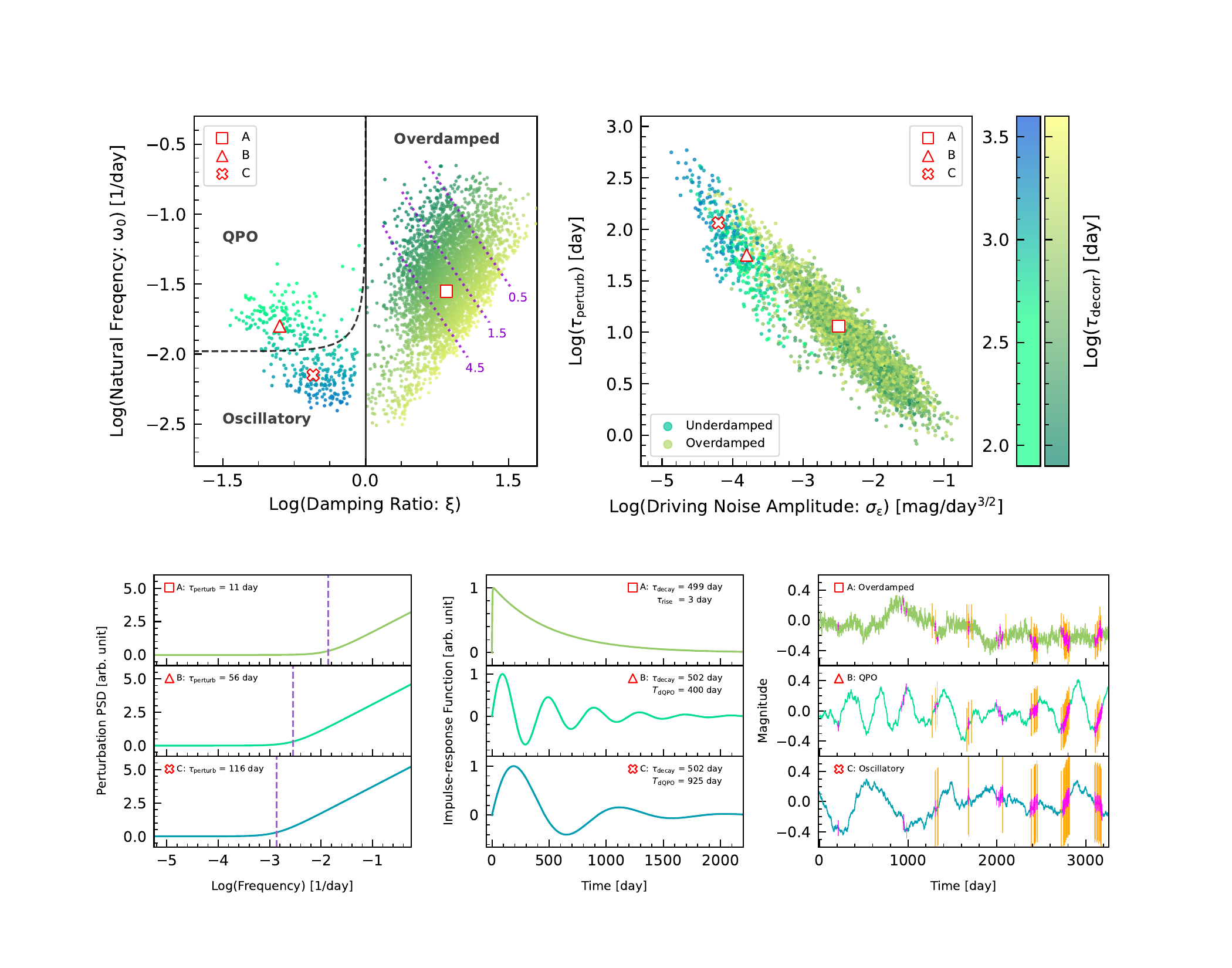}
    \vspace{-1.2cm}
    \caption{{\em Top left}: DHO fits of S82 quasars in the $\xi$, $\omega_0$ space (LHS of Equation~\ref{eqn:dho2}). Underdamped DHOs are located to the left of the solid vertical line ($\xi$ = 1) and overdamped DHOs are located to the right of this same line. The dash-dotted line further (empirically) separates underdamped DHOs into QPOs and oscillatory DHOs. Both underdamped and overdamped DHOs are colored by their decorrelation timescales (\taudecorr) but using two different color maps. The purple dotted lines on the underdamped population show the evolution of \taurise\ across the parameter space, where the numbers give the \taurise~in the unit of days.
    {\em Top right}: DHO fits of S82 quasars in the \sigmanoise, \tauperturb\ space (RHS of Equation~\ref{eqn:dho2}). Same as in the top left panel, color shows the corresponding \taudecorr\ for each object and different color maps are used for the underdamped and overdamped populations.
    {\em Bottom left}: The driving process (RHS of Equation~\ref{eqn:dho2}) PSDs of the three selected points \textbf{A, B} and \textbf{C} shown in the top two panels$^{\ref{perturb_psd}}$. 
    {\em Bottom center}: The response functions (LHS of Equation~\ref{eqn:dho2}) due to a delta function impulse of the dynamical systems described by the same selected points \textbf{A, B} and \textbf{C}. 
    {\em Bottom right}: DHO light curves simulated using \textbf{A, B} and \textbf{C} as the input model parameters. \editone{Superimposed are the versions downsamped using the best (magenta) and the worst (orange) cadences of our SDSS light curves (in terms of the number of observations, baseline duration, and photometric accuracy) for each classification. Median photometric uncertainties are used for error bars.}
    }
    \label{fig:best_fit_dist}
    \vspace{.05cm}
\end{figure*}

These four timescales (two for each class: \revise{underdamped and overdamped}) derived from the LHS of Equation~\ref{eqn:dho2} set the foundation for describing the response of the modeled dynamical system to a delta function impulse perturbation, whereas \tauperturb\ from the RHS of Equation~\ref{eqn:dho2} characterizes how the dynamical system is being driven/excited. In Figure~\ref{fig:best_fit_dist}, the bottom left panel provides the driving component (RHS of Equation~\ref{eqn:dho2}) power spectrum densities for three example DHO processes\footnote{\label{perturb_psd}\editone{We note that the driving component PSD presented here is not necessarily representative of that for the true underlying physical process, rather the features embedded (e.g., \tauperturb) in this PSD will inform us about important characteristics of the true physical model. We refer interested readers to \citet{jones1990}, \citet{brockwell2001}, and \citet{kasliwal2017} for further details on the analytic form of DHO's PSD.}}, the bottom middle panel shows the impulse-response functions (LHS of Equation~\ref{eqn:dho2}), and the bottom right panel plots the resulting time series (light curves).

On top of the intrinsic timescales, M19 also defines a decorrelation timescale, which characterizes the timescale at which the system becomes de-correlated \revise{from an earlier excitation} (forgets about its past self),
\begin{eqnarray}
    \revise{\textit{Underdamped}:} \;\tau_{\text {decorr }} &\approx& \frac{\pi}{2} T_{\mathrm{QPO}}, \nonumber\\
    \revise{\textit{Overdamped}:} \;\tau_{\text {decorr }} &\approx& \frac{\pi}{2}(\tau_{\text {rise}}+\tau_{\text {decay}}), 
\end{eqnarray}

Lastly, the asymptotic root-mean-square (RMS) amplitude of a DHO process ($\sigma_{\mathrm{DHO}}$), which is jointly determined by the parameters in the LHS and RHS of Equation~\ref{eqn:dho1}, can be computed using \citep{brockwell2001},
\begin{equation}\label{eqn:dho_amp}
    \sigma_{\mathrm{DHO}} = \sqrt{\frac{\beta_{1}^{2}\alpha_{2} + \beta_{0}^{2}}{2\alpha_{1}\alpha_{2}}} =\sigma_{\mathrm{\epsilon}}\sqrt{\frac{\omega_{0}^{2}\tau_{\mathrm{perturb}}^{2} + 1}{2\xi\omega_{0}^{3}}}.
\end{equation}

\subsection{DHOs as GPs}\label{subsec:carma_as_gp}
For a given time series (light curve), CARMA (DHO in this work) parameters are commonly extracted by maximum likelihood, where the likelihood function can be calculated through Kalman recursion in the ``state-space" of CARMA \citep{jones1990, brockwell2001, kelly2014, kasliwal2017}. Recently, \citet{foreman-mackey2017} introduced a new algorithm for performing fast Gaussian process (GP) modeling and suggested that the likelihood function of CARMA processes can be computed in $\mathcal{O}(NJ^{2})$ based on this new algorithm ($N$ is the number of data points in a time series and $J$ is the autoregressive order ($p$) of a CARMA model); this new algorithm is also up to 10 times faster than the Kalman recursion method. In this work, we adopt the algorithm introduced by \citet{foreman-mackey2017} and express DHO processes as a special class of GPs. 

Although rarely recognized, a CARMA process driven by Gaussian noise (i.e., $\epsilon(t)$ is Gaussian) is a Gaussian process, which makes it viable to calculate the likelihood function of CARMA models using GPs. The speedup demonstrated in \citet{foreman-mackey2017} originates from the fact that the CARMA auto-covariance function can be formulated as a sum of complex exponentials allowing the covariance matrix to be semi-separable, thus enabling faster computation of the likelihood function~\citep{ambikasaran2015, foreman-mackey2017}. Below we demonstrate how to represent a DHO's auto-covariance function \editone{(or entries in the auto-covariance matrix for discretely sampled data)} in terms of \texttt{celerite} kernels---the actual implementation of the algorithm presented in \citet{foreman-mackey2017}. The full derivation for CARMA processes of all orders is beyond the scope of this work and can be found in (Yu et al. 2022, in preparation).

From Equation~4 in \citet{kelly2014}, we can write out the auto-covariance function of a DHO process, 
\begin{equation}\label{eqn:dho_acvf1}
    R(\tau) = A_{1}e^{r_{1}\tau} + A_{2}e^{r_{2}\tau},
\end{equation}
\vspace{-.5cm}
\begin{eqnarray}\label{eqn:dho_acvf2}
    A_{1} = \frac{(\beta_{0}+\beta_{1}r_{1})(\beta_{0}-\beta_{1}r_{1})}{-2\,\mathrm{Re}(r_{1})*(r_{2}-r_{1})(r_{2}^{*}+r_{1})}, \nonumber\\
    A_{2} = \frac{(\beta_{0}+\beta_{1}r_{2})(\beta_{0}-\beta_{1}r_{2})}{-2\,\mathrm{Re}(r_{2})*(r_{1}-r_{2})(r_{1}^{*}+r_{2})},
\end{eqnarray}
where $r_{1}, r_{2}$ are the two roots of the characteristic polynomial associated with the LHS of Equation~\ref{eqn:dho1} \editone{and $\tau$ is the positive time lag between any two timestamps}.
Note that 
\citet{kelly2014} factored out $\beta_{0}$ and \revise{called} it $\sigma$---equivalent to \sigmanoise\ defined in the previous section.

When $r_{1}, r_{2}$ are real (overdamped DHOs: $\xi > 1$), $\mathrm{A}_{1}, \mathrm{A}_{2}$ are also real, and Equation~\ref{eqn:dho_acvf1} becomes a sum of two real exponential GP kernels (\texttt{celerite} real term). In the \texttt{celerite} framework, Equation \ref{eqn:dho_acvf1} for overdamped DHOs can be written as:
\begin{equation}\label{eqn:dho_real_kernel}
    k \left(\tau_{n m}\right)=\sigma_{n}^{2} \delta_{n m} + a_{1}\,e^ {-c_{1}\tau_{n m}} + a_{2}\,e^{-c_{2}\tau_{n m}}
\end{equation}
where $\tau_{nm}$ is the \editone{positive time lag} between the $n^{th}$ and $m^{th}$ data point in a time series, $\sigma_{n}$ is the measurement uncertainty on the $n^{th}$ data point; $a_{1}=A_{1}$, $a_{2}=A_{2}$, $c_{1}=-r_{1}$, and $c_{2}=-r_{2}$. \editone{Note that for a discretely sampled DHO process, $k\,(\tau_{n m})$ gives the entries in the corresponding auto-covariance matrix.}

In the case of underdamped DHOs ($\xi < 1$), $r_{1}, r_{2}$ are complex conjugates, \revise{so are $A_{1}$, $A_{2}$}. Equation~\ref{eqn:dho_acvf1} becomes a single complex exponential kernel:
\begin{eqnarray}\label{eqn:dho_complex_kernel}
k \left(\tau_{nm}\right) =&&\sigma_{n}^{2} \delta_{n m} + \frac{1}{2}\left(a+i b\right) e^ {-\left(c+i d\right) \tau_{nm}} \nonumber\\
&&+\frac{1}{2}\left(a-i b\right) e^ {-\left(c-i d\right) \tau_{nm}}
\end{eqnarray}
where $a = 2*\mathrm{Re}(A_{1})$, $b = 2*\mathrm{Im}(A_{1})$, $c = -\mathrm{Re}(r_{1})$, and $d = -\mathrm{Im}(r_{1})$. 
Given this mapping from \revise{a} DHO's auto-covariance function to \texttt{celerite}'s GP kernels, we can take advantage \texttt{celerite} to fit DHO to our light curves. 

\subsection{Fitting DHO to Quasar Light Curves}\label{subsec:fit_dho}

In the previous section, we demonstrated that the \texttt{celerite} framework can be utilized to compute the likelihood function of CARMA (used DHO as an example), however, the native \texttt{celerite} software does not come with the functionality (the actual code) to fit an arbitrary CARMA model that is more complex than the DRW to light curves. 
To facilitate general-purpose CARMA modeling taking advantage of \texttt{celerite}, we implemented the appropriate mapping from the CARMA parameterization to the \texttt{celerite} parameterization in a new Python package---\texttt{EzTao}, which was used to fit DHO to our quasar light curves.

The likelihood landscape of a complex GP kernel (such as that of a DHO) is \revise{usually} non-convex (e.g., having multiple local optima).
Thus, during the fitting process we randomly initialized 100 optimizers across the DHO parameter space that can be probed by the temporal sampling of the input light curves and selected the maximum {\em a posterior} (MAP) estimation as the best-fit solution, where very broad flat priors were used to prevent the potential numerical overflow/underflow caused by catastrophic runaways of the optimizers\footnote{The AR coefficients ($\alpha_1$ and $\alpha_2$ from the LHS of Equation~\ref{eqn:dho1}) have a boundary of [-15, 15] in the natural log scale, and the MA coefficients ($\beta_0$ and $\beta_1$ from the RHS of Equation~\ref{eqn:dho1}) have a boundary of [-23, 7] in the natural log scale}.
We performed the fitting process 5 times for each light curve to make sure that a robust fit was obtained. Our experience showed that neither increasing the number of optimizers nor repetitions will change the final distribution of the best-fit DHO parameters for our quasar sample. 

After we obtained the best fit for each object, we used \texttt{emcee} \citep{foreman-mackey2013}, a python implementation of Goodman \& Weare’s Affine Invariant Markov chain Monte Carlo (MCMC) Ensemble sampler \citep{goodman2010}, to sample the posterior distribution with the MCMC walkers initiated at the MAP position. 
\editone{One additional prior was used to restrict the MCMC walkers from potential catastrophic runaways: $10^{-3}$ days $<$ log$\,(\tau_{\mathrm{perturb}})$ $<$ $10^{5}$ days. We ran MCMC for 15,000 steps using 32 walkers and discarded the first 3,000 steps as the ``burn-in". The largest auto-correlation time for all chains is around 500 steps. The uncertainty of the best-fit DHO parameter is taken as the ``1-sigma" range (one half the central 68.3\% interval) of the marginalized posterior distribution.}

\section{Best-fit DHO Parameters}\label{sec:best_fit}

\subsection{The Distribution of DHO Parameters and The Identification of Bad Fits}\label{subsec:best_fit_dist}

\begin{figure}[ht!]
    \vspace{-.8cm}
    \centering
    \includegraphics[width=0.42\textwidth]{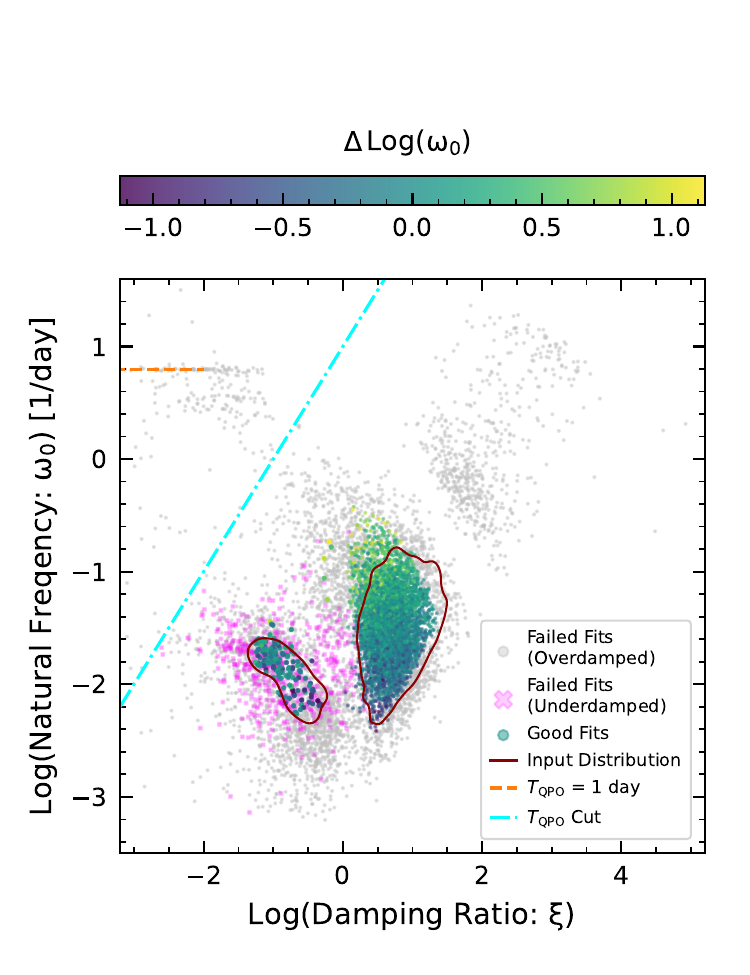}
    \vspace{-.3cm}
    \caption{
    DHO fits for realistically simulated SDSS S82 light curves, the input model parameters are drawn from the distribution shown in the top two panels of Figure~\ref{fig:best_fit_dist}. 
    The DHO fits that are identified as good using the algorithm described at the beginning of Section~\ref{subsec:best_fit_dist} are color-coded by the difference between their best-fit log($\omega_{0}$) and the input. 
    \revise{Identified bad fits that were simulated as overdamped DHOs are shown using gray dots and those simulated as underdamped DHOs are shown using magenta crosses.}
    The orange dashed line marks a $T_{\mathrm{QPO}}$ of one day. The cyan dash-dotted line shows the cut used to remove the bad fits associated the daily observing cadence (surrounding the orange dashed line). The red contour outlines \revise{the distribution of the input} DHO parameters.}
    \label{fig:fits_on_sim_dist}
\end{figure}

Just because a fit is robust (in that it does not change across multiple independent runs) does not mean that it is accurate and precise. 
We identified two main ``dead zones" in the DHO parameter space that are hosts of bad DHO fits, these are MAP fits that are catastrophic failures potentially as a result of insufficient temporal sampling or large photometric uncertainty. 
DHO fits that end up in those regions, as listed below, were flagged as bad and removed from our sample:
\begin{enumerate}\label{list:quality_cut}
    \item DHO fits corresponding to timescales (see Section~\ref{subsec:dho}) that are either longer than the span of the light curve (\maxdt)\footnote{\editone{This cut will bias the distribution of \taudecorr, which has a median of 1000 days in the observed-frame---comparable to the baseline of our light curves. However, we stress that such a cut is needed since an intrinsic decorrelation timescale longer than the light curve span can not be constrained. See \citet{kozlowski2017a, kozlowski2021} for close examinations of this effect on the DRW model.}} or shorter than one half the minimum separation between any two observations (\mindt).
    \item DHO fits with log($\xi$) -- log($\omega_{0}$) $>$ 1; for those objects, the daily observing cadence exhibits a stronger signal than the intrinsic timescales (see Figure~\ref{fig:fits_on_sim_dist}).
\end{enumerate}

\editone{The first cut above alone removed $\approx$70\% of the objects from our initial sample in every photometric band. Independently from the first cut, the second cut alone removed $\approx$55\% of all objects in $u$ and $z$, and $\approx$20\% of all objects in $gri$. After these two cuts, we were left with 2997, 4570, 3530, 2999, 1180 DHO fits in $ugriz$ bands, respectively.}
\revise{Next}, we used an isolation forest (an outlier detection algorithm) \citep{liu2008} to further identify and remove bad fits. An isolation forest takes a data set and splits it randomly until no more split can be made. If anomalies/outliers are rare and different from the main population, it would take longer (more splits) to isolate a ``regular" data point than an outlier. Therefore, those that are first isolated are identified as outliers by an isolation forest. 
\editone{Running an isolation forest on the current sample of DHO fits in each band separately removed an additional 5\% of objects,} our final sample contains 2847, 4341, 3353, 2849, 1121 good DHO fits in $u, g, r, i, z$ bands, respectively. 

The distribution of good DHO fits (using $g$-band as an example) is shown in the top two panels of Figure~\ref{fig:best_fit_dist}: 
the top left panel shows the distribution in the $\xi$, $\omega_{0}$ space (the response component, LHS of Equation~\ref{eqn:dho2}) and the top right panel shows that in the $\sigma_{\mathrm{\epsilon}}$, $\tau_{\mathrm{perturb}}$ space (the driving perturbation component, RHS of Equation~\ref{eqn:dho2}).
In the top left panel of Figure~\ref{fig:best_fit_dist} we can identify two main clusters, one for underdamped DHOs (log$(\xi) < 0$) and one for overdamped DHOs (log$(\xi) > 0$), separated by the vertical solid line at log$(\xi) = 0$. The clustering of underdamped/overdamped populations is not a characteristic of the DHO model rather that of the quasar light curves, that is, any point in the DHO parameter space constitutes a valid model. 
We can further classify underdamped DHOs into quasi-periodic (QPO) and oscillatory DHOs with a (empirically chosen) dividing $T_{\mathrm{dQPO}}$ of 600 days (dashed line). \editone{Despite both being underdamped, QPOs generally have larger natural oscillation frequencies ($\omega_0$) and smaller damping ratios than oscillatory DHOs. }Both underdamped and overdamped DHOs are colored by their \taudecorr~but using different color maps. The dotted purple lines on top of the overdamped population demonstrate the evolution of \taurise\ across the response parameter space, where the numbers give the \taurise~in the unit of days.
In the top right panel (the perturbation parameter space), objects are also colored by \taudecorr. Since we do not see an obvious color gradient as \tauperturb~increases (or decreases) \revise{in the overdamped cluster}, we argue that $\tau_{\mathrm{decorr}}$ and $\tau_{\mathrm{perturb}}$ are largely uncorrelated \revise{for overdamped DHOs}, which suggests that the perturbation component (RHS of Equation~\ref{eqn:dho2}) and the response component (LHS of Equation~\ref{eqn:dho2}) of the modeled quasars might be decoupled. Such decoupling is expected if the perturbation process is external to instead of originating in the accretion disk of AGN (see Section~\ref{sec:discussion} for further discussion).

Three representative points are chosen from the distribution of good DHO fits, one for each identified class: overdamped DHO, QPO, and oscillatory DHO, to investigate/visualize the intrinsic variability signatures embedded in the DHO parameters; they are labeled using a square (overdamped DHO), triangle (QPO), and a cross (oscillatory DHO). 
The bottom three panels of Figure~\ref{fig:best_fit_dist}, from left to right, show the perturbation spectrum---the driving process PSD (RHS of Equation~\ref{eqn:dho2}), the impulse-response function (LHS of Equation~\ref{eqn:dho2}), and the final (simulated) DHO light curves, of the dynamical systems described by the corresponding DHO parameters at the selected points. 

It can be clearly seen from these three example DHOs that different regions of the parameter space correspond to different unique characteristics. The underdamped population---log($\xi$) $< 0$ in the top left panel of Figure~\ref{fig:best_fit_dist} and the top-left corner in the top right panel of Figure~\ref{fig:best_fit_dist}---has large \tauperturb\ and the light curve is smooth at short timescales but bumpy at long timescales; 
\editone{comparatively, the QPO subclass shows a much stronger periodicity than the oscillatory subclass.}
On the other hand, the overdamped population---log($\xi$) $> 0$ in the top left panel of Figure~\ref{fig:best_fit_dist} and the bottom-right portion in the top right panel of Figure~\ref{fig:best_fit_dist}---has smaller \tauperturb\ and the light curve is bumpy at short timescales but smooth over large timescales.

Next, to confirm that our bad-fit identification procedure works as expected, we simulated $\sim$12,000 DHO light curves sampled at the exact temporal cadence and photometric accuracy of the quasars in our \textbf{main sample}, and then fitted them with DHO. The input model parameters for the simulated light curves were drawn from the distribution of $g$-band good fits (see top two panels Figure~\ref{fig:best_fit_dist}).
Figure~\ref{fig:fits_on_sim_dist} shows the distribution of the MAP best fits obtained on these simulated data in the $\xi$, $\omega_{0}$ parameter space.
Based on the quality cuts described at the beginning of this section, DHO fits that passed the cuts are color-coded by the difference in log($\omega_{0}$) between the MAP value and the input; the simulated overdamped DHOs that failed to pass the cuts are shown using gray dots whereas the failed underdamped DHOs are shown using magenta crosses.
We can see that the distribution of the identified good fits largely overlaps that of their input outlined by the red contour. 
We can also see that the bad fits that were simulated as overdamped DHOs spread over the whole $\xi$, $\omega_{0}$ parameter space and the bad fits that were simulated as underdamped DHOs are more concentrated around the distribution of their input.
The orange dashed line at the top left corner corresponds to a $T_{\mathrm{QPO}}$ of one day suggesting that best fits accumulated in this region might be fitting the daily observing cadence rather than the intrinsic timescales in the quasar light curves. A cut of log($\xi$) -- log($\omega_{0}$) $>$ 1 (the cyan dash-dotted line in Figure~\ref{fig:fits_on_sim_dist}) effectively removes those suspicious fits.
Given the resulting distribution of good/bad fits obtained on simulated light curves, we can confirm that the good fits identified through the criteria established at the start of this section are effective. However, we do find two flavors of overestimation/underestimation from the distribution of good fits: a systematic offset (between the colored distribution and the red contour) and a parameter-dependent trend (the color gradient within the colored distribution); we will discuss both scenarios with more detail in the next section.

\subsection{The Effects of Time Sampling and Photometric Accuracy on Best-fit Parameters and Possible Corrections}\label{subsec:cadence_effect}
\begin{figure*}[ht!]
    \centering
    \vspace{0.2cm}
    \includegraphics[width=0.95\textwidth]{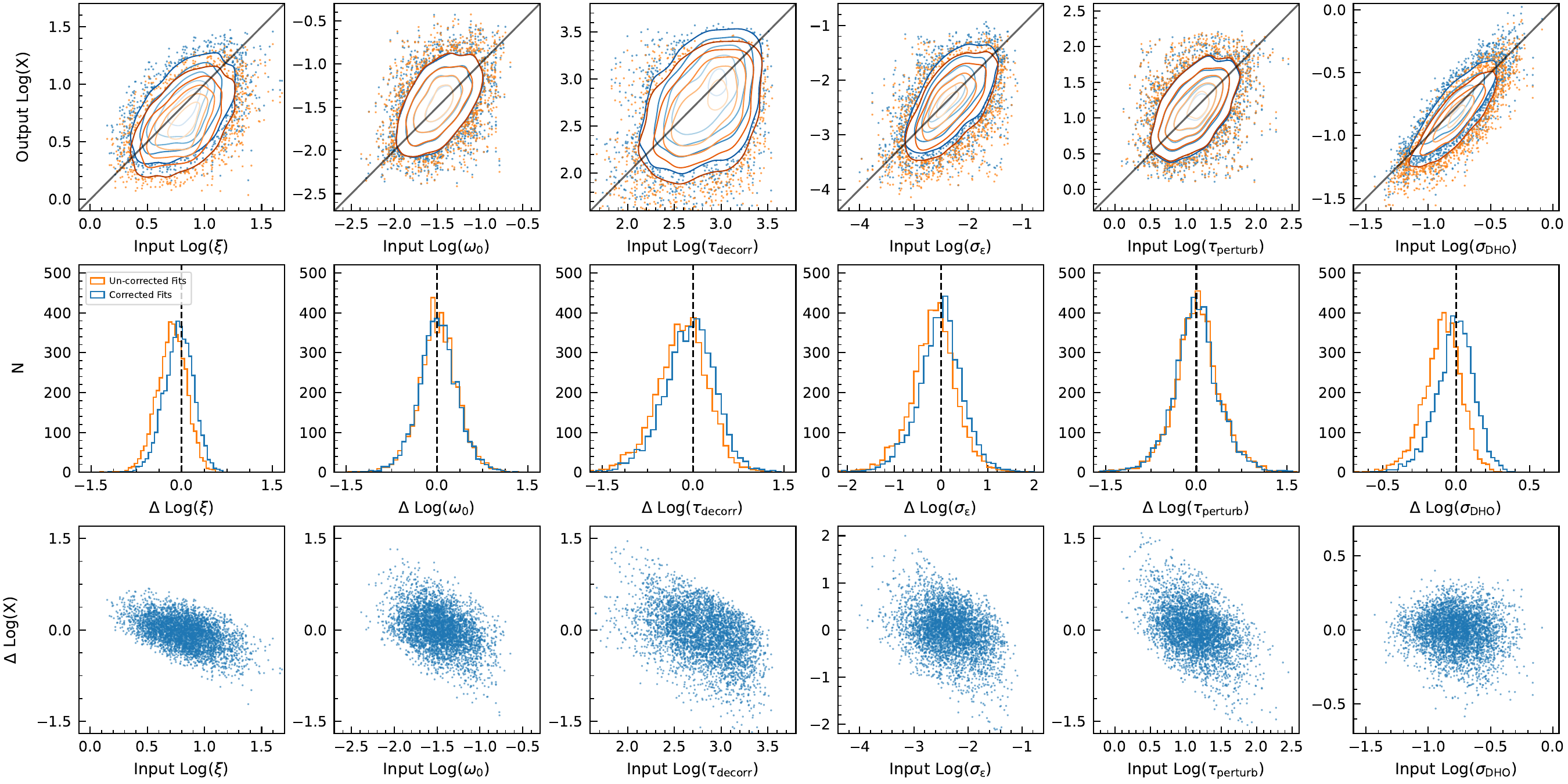}
    \vspace{0.15cm}
    \caption{A comparison between the output DHO parameters obtained on simulated light curves and the true input; only simulations of overdamped DHOs are shown. Two versions of the output parameters are shown: the first one (orange) is the original MAP best fit and the second (blue) has the linear correction applied. The correction is determined from modeling the difference between the output and \revise{the} input parameters in the form of $\Delta \,$log(X) = log($\mathrm{X_{Output}}$) - log($\mathrm{X_{Input}}$), as a function of the temporal sampling and photometric accuracy of the light curves.
    {\em Row One}: Scatter plots comparing \revise{the} output DHO parameters (y-axis) with the input parameters (x-axis). Solid diagonal lines show a one-to-one correspondence. 
    {\em Row Two}: Histograms showing the difference between the output DHO parameters and \revise{the input}. The vertical dashed lines mark an offset of zero meaning the output is in perfect agreement with the input. 
    {\em Row Three}: The difference between the (corrected) output parameters and the input parameters as a function of the input parameters. 
    }
    \label{fig:fits_sim_correction}
\end{figure*}

It has been previously reported that the accuracy/precision of the maximum likelihood estimate of DRW parameters is sensitive to the ``quality" of the light curves, more specifically, the ratio between the DRW decorrelation timescale ($\tau_{\mathrm{DRW}}$) and the light curve length (\maxdt) \citep{kozlowski2017a, kozlowski2021}.
DHO and DRW are the same class of stochastic diffusion processes---CARMA, thus we expect similar trends to appear for best-fit DHO parameters determined by maximum likelihood (or maximum {\em a posterior} with wide flat priors). 

The light curve ``quality" measurements used in \citep{kozlowski2017a} depend on both the true parameters of the underlying process (e.g., $\tau_{\mathrm{DRW}}$) and some basic properties of the light curve data (e.g., \maxdt), where the former are hardly known but the latter are easily measurable.
Thus, rather than trying to fully characterize the accuracy/precision of our best-fit DHO parameters with respect to those ``quality" metrics, we attempted to calibrate the best-fit DHO parameters using simulated data. The same simulation process as described at the end of Section~\ref{subsec:best_fit_dist} was carried out for all five bands, and the best-fit parameters from different bands were joint together to determine the correction.
We \revise{first} modeled the offset of the best-fit DHO parameters on simulated light curves relative to the input parameters (log$(\mathrm{X_{Output}}) - $log$(\mathrm{X_{Input}})$) as a multivariate linear function of three basic properties of the light curves: the total length of the light curve (\maxdt), the minimum separation between any two observations (\mindt), and the median photometric uncertainty. 
\editone{The coefficients of the best-fit linear regression suggest that the offset is most correlated with the median photometric uncertainty and \maxdt---consistent with the results from previous work for DRW~\citep{kozlowski2017a}.}

By applying a correction derived from the best-fit multivariate regression, we were able to remove the systematic offset between the input and \revise{the} output distribution of DHO parameters. The top six panels of Figure~\ref{fig:fits_sim_correction} show the comparison between the input parameters and the output (MAP best-fit) parameters before (orange) and after (blue) the applied correction; the middle six panels show the distributions of the offsets (log$(\mathrm{X_{Output}}) - $log$(\mathrm{X_{Input}})$). The corrected fits have better correspondence with the input parameters than the original ones (i.e., the blue histograms in the second row are more centered at zero than the orange ones). The panels on the third row of Figure~\ref{fig:fits_sim_correction} show the offsets of the corrected best fits as a function of the input parameters; any parameter-dependent trends as shown were not accounted for in our correction and are left for future investigations. From simulations, we can see that given the light-curve cadence and photometric accuracy of S82 quasars, \sigmadho, \tauperturb, and \sigmanoise~are best constrained among all other DHO features---in terms of the size of \revise{the} dispersion and \revise{the} level of the parameter-dependent trend (modulo the extremes) of the offsets. 


In addition to the overestimation/underestimation of DHO parameters, it would be interesting to investigate how reliable is the DHO subclass classification (underdamped DHO vs.~overdamped DHO) given the light-curve cadence and photometric accuracy of S82 quasars. Using the simulated data introduced above, we computed the classification \revise{precision} (and recall) for both populations. 
Given a clean sample, for a particular classification, the \revise{precision} is defined by the percentage of best-fit DHOs that are also simulated with the same classification (e.g., a best-fit overdamped DHO is also simulated as an overdamped DHO) and the recall is defined by the percentage of simulated DHOs that are correctly classified. The final result is shown in Table~\ref{tab:dho_class}. Since the classification of underdamped DHOs into QPOs and oscillatory DHOs is based on an empirically chosen diving $T_{\mathrm{dQPO}}$, it is not as informative to show the statistics for those two classes.

\begin{table}[h!]
\caption{DHO Classification \revise{Precision} (Recall)}
\hspace{-1.cm}
\centering
\begin{tabular}{cll}
\hline\hline
band  &  Underdamped DHO    & Overdamped DHO\\
\hline
u       &    14.29 (100.00)  &  100.00 (90.65)   \\
g       &    29.80 (100.00)  &  100.00 (97.56)   \\
r       &    29.46 (92.68)   &  \, 99.93 (97.90) \\
i       &    24.16 (100.00)  &  100.00 (96.58)   \\
z       &    \, 9.18 (100.00)&  100.00 (87.85)   \\
\hline
\end{tabular}
\vspace{-.4cm}
\label{tab:dho_class}
\end{table}


In summary, given the temporal sampling and photometric accuracy of SDSS Stripe 82 quasar light curves, more than 99.9\% of the overdamped DHOs have the correct classification across all five SDSS bands. Among those classified as underdamped DHOs, the \revise{precision} is between $\approx$9\% and $\approx$29\% with the highest in the $g$ and $r$ bands and the lowest in the $z$ band. The recall for the overdamped population is between $\approx$88\% and $\approx$98\% with the highest in the $g$ and $r$ bands and the lowest in the $z$ band, and for the underdamped population all bands have a recall of 100\% except for the $r$ band (92.68\%).
Since failed overdamped DHO fits can end up in the region of the underdamped population and failed underdamped DHO fits tend to stay close to their input (see grey dots and magenta crosses in Figure~\ref{fig:fits_on_sim_dist}), we suspect that the low classification \revise{precision} of the underdamped population is partially caused by the overabundance of the overdamped population. That said, given that our input parameters were drawn from the distribution of MAP best fits obtained on real SDSS Stripe 82 quasar light curves, the true relative abundance and classification \revise{precision} of underdamped DHOs should be smaller than what we are showing in Figure~\ref{fig:best_fit_dist} and Table~\ref{tab:dho_class}.

\section{Correlations of DHO Parameters with Wavelength and Physical Properties of AGN} \label{sec:results}

We investigated the correlations between the best-fit DHO parameters (corrected using the linear coefficients determined in Section~\ref{subsec:cadence_effect}) and the \editone{estimated} physical properties of the quasars in our sample \citep{shen2011, rankine2020}. 
Our investigation was focused on the overdamped population provided that the classification for underdamped DHOs is highly unreliable (see Table~\ref{tab:dho_class}).
We also limited our analysis to three DHO features: $\sigma_{\mathrm{DHO}}$, $\sigma_{\mathrm{\epsilon}}$ and $\tau_{\mathrm{perturb}}$, because they are best constrained given the light-curve cadence and photometric accuracy of SDSS Stripe 82 quasars (see Figure~\ref{fig:fits_sim_correction}) and that they are least affected by our initial timescale-based cut for removing bad DHO fits (the first criterion listed at the start of Section~\ref{subsec:best_fit_dist}).
More specifically, the distribution of \tauperturb\ lies between \mindt\ and \maxdt, thus, is not affected by the cuts associated with those two timescales. On the other hand, \taurise\ is on the scale of \mindt\ ($\approx$ 1 day) and \taudecay\ is on the scale of $\frac{1}{2}\,\mathrm{Max}\,\Delta t$ ($\approx$ 1000 days), therefore the distribution of \taurise\ and \taudecay\ are subject to selection bias, and similarly for \taudecorr $\,\approx \frac{\pi}{2}$(\taurise\ + \taudecay). 
Lastly, through examining the MCMC samples of our good DHO fits we found that some of them have bi/multi-modal and/or very broad posterior distributions (see Appendix~\ref{apx:bimodal_mcmc} for examples). We suspect this to be caused by the irregular/sparse sampling of S82 light curves that leaves certain timescales less well probed and/or the (lack of) variability of strong emission line(s) in the particular photometric bands. Further investigations are needed to verify the origin(s) of those bi/multi-modal posterior distributions. Nevertheless, we removed those objects from the following analysis using a well-defined metric (see Appendix~\ref{apx:bimodal_mcmc}). 


\subsection{Wavelength Dependence of DHO Parameters}\label{subsec:lambda_dep}

Given the large redshift range that our quasar sample spans and the five different photometric bands that we are utilizing in this investigation, best-fit DHO parameters should first be evaluated against and corrected for any wavelength dependence before being used to correlate with the physical properties of quasars. We explored the wavelength dependence of \sigmadho, \tauperturb, and \sigmanoise~by first correcting them for redshift dependence and then plotting them as a function of the effective wavelength of their photometric bands in rest frame~\citep{schneider1983}. \sigmadho, \tauperturb, and \sigmanoise\ scale with redshift as $(1+z)^{0}$, $(1+z)^{-1}$, and $(1+z)^{3/2}$, respectively; the derivation for the redshift dependence of \sigmanoise~can be found in Appendix~\ref{apx:b0_z}. The effective wavelength, computed based on a power-law continuum with a spectral index $\alpha_v$ of $-0.5$, of the SDSS $ugriz$ bands in the observer's frame are 3541, 4653, 6147, 7461, and 8904 angstroms, respectively \citep{fukugita1996, richards2001, kaczmarczik2009}.

\begin{figure}[h!]
    \centering
    \includegraphics[width=0.44\textwidth]{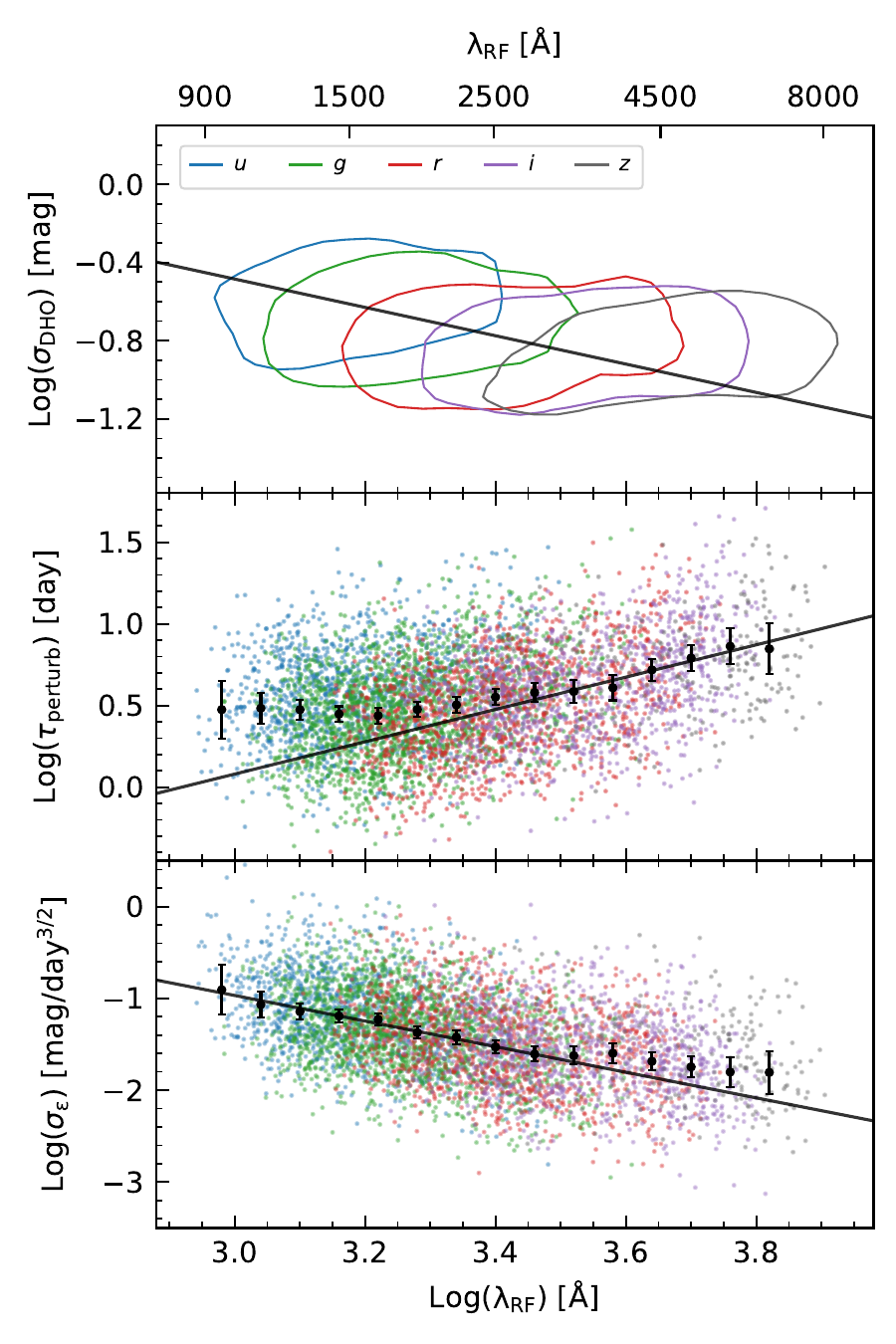}
    \caption{Distribution of \sigmadho, \tauperturb, and \sigmanoise~as a function of rest wavelength (\lambdarf). The color gives the photometric band from which the best-fit parameters are obtained. The black points in the middle and bottom panels mark the median values in bins centered at the corresponding \lambdarf; the error bars have been made twice as large for better visibility.
    {\em Top}: The contours show the 85\% mass for each band. \sigmadho~follows an overall monotonic relation with \lambdarf; the solid line shows the best linear fit. {\em Middle}: \tauperturb~increases with \lambdarf~longward of $\approx$2500 \AA; the solid line shows the best linear fit to the median values with \lambdarf~$>$ 2500 \AA. 
    {Bottom}: \sigmanoise~decreases with \lambdarf~shortward of $\approx$2500 \AA; the solid line shows the best linear fit to the median values with \lambdarf~$<$ 2500 \AA.}
    \label{fig:lambda_dp_all_scatter}
\end{figure}

Figure~\ref{fig:lambda_dp_all_scatter} shows \sigmadho, \tauperturb, and \sigmanoise\ as a function of the effective wavelength of their photometric bands in rest frame (\lambdarf), the color indicates the specific passband from which the best-fit parameters are obtained. \lambdarf\ for a given band is computed as: \lambdarf\ = $\lambda_{\mathrm{eff}}/(1+z)$, where $\lambda_{\mathrm{eff}}$ is the effective wavelength in the observer's frame and $z$ is the redshift of an object. 
From the top panel we can see that \sigmadho\ follows an overall monotonically decreasing trend with \lambdarf---in agreement with previous findings \citep[e.g.,][]{vandenberk2004, macleod2010}. 
We performed a bisector linear regression to derive the best-fit relation between \sigmadho~and \lambdarf,
\begin{eqnarray}\label{eqn:amp_lambda}
    \mathrm{log}(\sigma_{\mathrm{DHO}}) = (\editone{-0.73}~&&\pm~\editone{0.070})~\mathrm{log}(\lambda_{\mathrm{RF}})\nonumber\\ 
    +\editone{1.69}~&&\pm~0.003.
\end{eqnarray}
Note that \sigmadho\ follows an increasing rather than a decreasing trend with \lambdarf\ within each individual band. 

We suspect this ``misleading" increasing trend to be caused by the anti-correlation of \sigmadho\ with \lbol~(see Section~\ref{subsec:result_amp}) and the selection bias intrinsic to flux-limited samples (i.e., selected objects from higher redshift are systematically more luminous than those from lower redshift). 

The middle panel of Figure~\ref{fig:lambda_dp_all_scatter} shows that \tauperturb~(the characteristic timescale of the perturbation component) increases with rest wavelength at \lambdarf~$>$ 2500~\AA~with a best-fit relation of \editone{log(\tauperturb) = $0.99*$log(\lambdarf) $-\,2.89$}, but is nearly independent of wavelength at \lambdarf~$<$ 2500~\AA. 

In the bottom panel of Figure~\ref{fig:lambda_dp_all_scatter}, we see that \sigmanoise~(the amplitude of the driving white noise) decreases with an increasing \lambdarf~until $\approx$2500 \AA~and then becomes nearly independent (or only weakly dependent) of wavelength. At \lambdarf~shorter than 2500~$\mathrm{\AA}$, the best-fit relation between \sigmanoise~and \lambdarf~is \editone{log(\sigmanoise) = $-1.39*$log(\lambdarf) $+\,3.21$}.

We note that the trends of \sigmadho, \tauperturb, and \sigmanoise\ with \lambdarf\ shown in Figure~\ref{fig:lambda_dp_all_scatter} persist when we replace the corrected DHO parameters with their un-corrected version, however, the v-shaped correlation shown in the bottom panel for \sigmanoise\ will appear less obvious. 

Both \tauperturb~and \sigmanoise~exhibit clearly different dependencies with wavelength on either side of 2500 \AA~while \sigmadho~only shows a largely monotonic dependency. We will further evaluate this interesting feature in Section~\ref{subsec:warm_disk} and discuss what could be implied regarding accretion disk models.

\subsection{DHO Amplitude}\label{subsec:result_amp}

\begin{figure}
    \centering
        \includegraphics[width=0.47\textwidth]{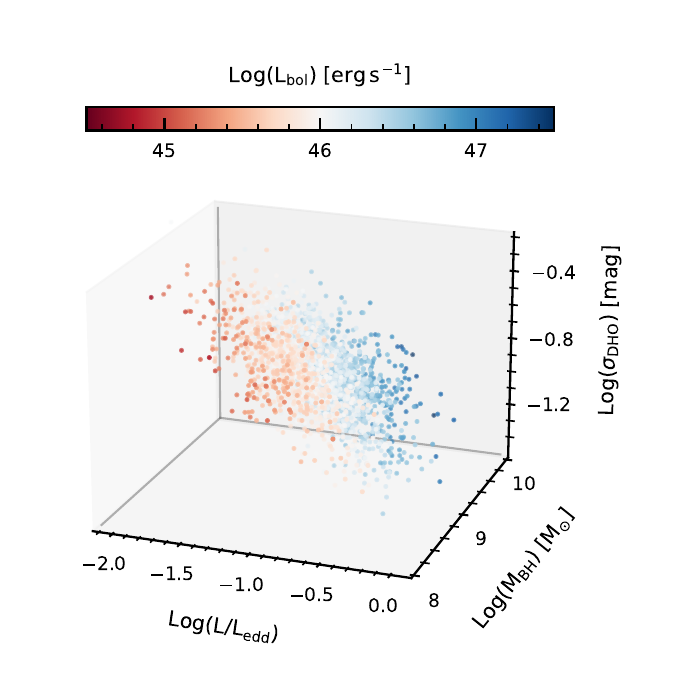} 
    \caption{A 3D plot demonstrating the anti-correlations of \sigmadho\ with \lledd\ and \mbh. The x-axis is \lledd, the y-axis is \mbh, and the z-axis is \sigmadho. The color indicates \lbol\ for each object. An interactive version of this figure is available online.}
    \label{fig:amp_edd_ratio}
\end{figure}

The variability amplitude has been studied most extensively among all other variability metrics in terms of searching for correlations with the fundamental properties of AGN (e.g., luminosity) \citep{vandenberk2004, wilhite2007, macleod2010, zuo2012}. 
It is worthwhile to check if $\sigma_{\mathrm{DHO}}$ exhibits similar correlations with those fundamental properties of the quasars in our sample. In this part of our analysis, \sigmadho~is corrected to a rest-frame wavelength of 2500 \AA\ using Equation~\ref{eqn:amp_lambda}, and for each quasar only the DHO fit from the photometric band with the smallest uncertainty in \sigmadho~(determined from MCMC) is used. We also limit our quasars to $0.7 < z < 1.9$ for the sake of enforcing a consistent \revise{determination} of \lbol, \mbh, and \lledd~\citep[i.e., $L_{\mathrm{3000}}$ was used to \editone{estimate} \lbol\ and the Mg\,II emission line was used to \editone{estimate} \mbh;][]{shen2011}.
Indeed, we confirm that $\sigma_{\mathrm{DHO}}$ is anti-correlated with the \editone{recorded values of \lledd~and \mbh\ for our quasars}. 
Figure~\ref{fig:amp_edd_ratio} demonstrates said anti-correlations by displaying the distribution of our quasars in the space of \sigmadho, \lledd\, and \mbh.
We fitted a multivariate bisector regression to \sigmadho, \lledd, and \mbh~with the best-fit relation shown by Equation~\ref{eqn:amp_lledd}.
\begin{eqnarray}\label{eqn:amp_lledd}
    \mathrm{log}(\sigma_{\mathrm{DHO}}) = 
    (\editone{-0.41}~&&\pm~\editone{0.012})\,\mathrm{log}(\mathrm{L/L_{Edd}})\nonumber\\
    +(\editone{-0.30}~&&\pm~\editone{0.011})\,\mathrm{log(M_{BH})}\nonumber\\
    +(\editone{+1.48}~&&\pm~\editone{0.092})
\end{eqnarray}
Note this trend can also be expressed equivalently as an anti-correlation with \lbol~in addition to a relatively weaker positive correlation with \mbh~\citep{wilhite2007, macleod2010, zuo2012}.

\begin{figure*}[ht!]
    \centering
    \vspace{0.1cm}
    \includegraphics[width=0.4\textwidth]{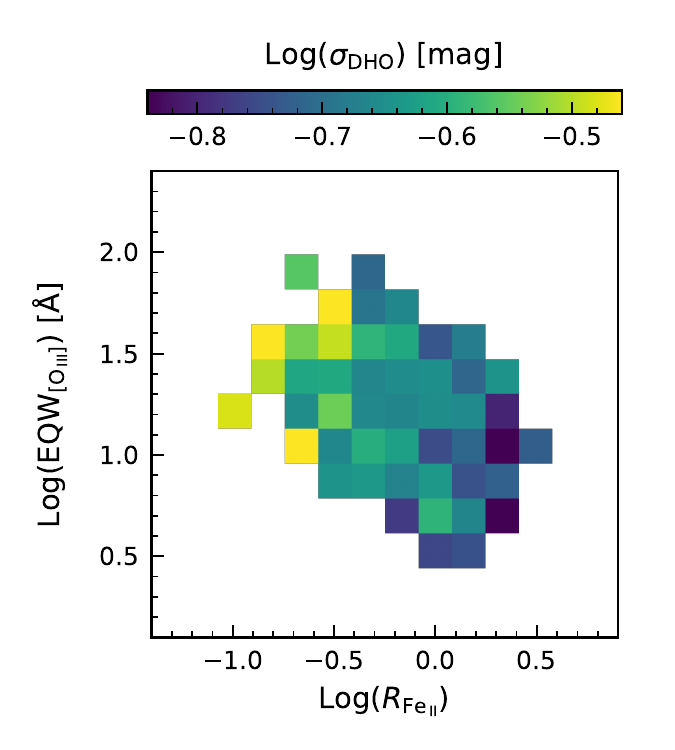}
    \hspace{0.1cm}
    \includegraphics[width=0.4\textwidth]{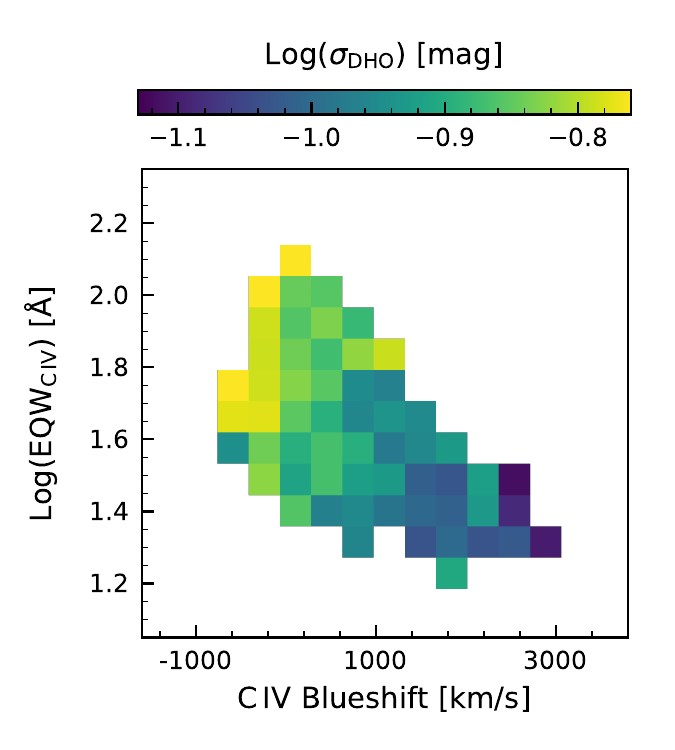}
    \caption{DHO amplitude (\sigmadho) and emission line properties.
    {\em Left}: Distribution of quasars in the EV1 parameter space with the color showing the mean \sigmadho~of objects in the corresponding bins; a trend of increasing \sigmadho~towards the top left corner is clearly shown.  
    {\em Right}: Distribution of quasars in the \CIV\ parameter space with the color showing the mean \sigmadho~of objects in the corresponding bins; a trend of increasing \sigmadho~with decreasing \CIV~blueshift and increasing \CIV~EQW is shown.
    }
    \label{fig:amp_eml}
    \vspace{0.1cm}
\end{figure*}


We further explored the correlations between DHO amplitude and emission line properties, more specifically with the ``Eigenvector 1" \citep[EV1;][]{Boroson2002} sequence and \CIV~equivalent width (EQW) and blueshift. EV1 refers to a dominant trend observed among mainly low redshift AGN: the stronger the broad \FeII~emission, the weaker the narrow \OIII~emission. Following the convention established in \citet{shen2014}, we define the strength of the broad \FeII~(4434-4684~\AA) emission as the ratio between the broad \FeII~EQW and $\mathrm{H\beta}$ EQW, $R_{\mathrm{Fe\,II}} = \mathrm{EQW_{Fe\,II}}/\mathrm{EQW_{H\beta}}$; the \OIII~strength is characterized by its equivalent width. 
The left panel of Figure~\ref{fig:amp_eml} shows the distribution of our quasars in the EV1 parameter space; we binned our sample onto a uniform grid with the color indicating the average DHO amplitude. 
In addition to the expected anti-correlation between $\mathrm{EQW_{[O\,III]}}$ and $R_{\mathrm{Fe\,II}}$, we see a trend of increasing \sigmadho~toward the top left corner of this plot---consistent with the result from an earlier investigation conducted using a smaller sample \citep{ai2010}. The EV1 sequence has long been argued to be driven by the diversity in \lledd~\citep{Boroson2002, shen2014}, that is, data points at the top left corner should have smaller \lledd~than those at the bottom right corner. Indeed, this argument is \revise{consistent} with that implied by the anti-correlation between \sigmadho~and \lledd. 

At high redshift, \CIV~EQW and \CIV~blueshift alone are proposed to be indicators of \lledd~\citep{Shemmer2015, rankine2020}. The right panel of Figure~\ref{fig:amp_eml} shows the distribution of our sample in the \CIV\ parameter space; we used the same technique as utilized in the EV1 analysis to bin our data. 
We found that our quasars tend to have larger variability amplitude than average when \CIV~EQW is large and \CIV~blueshift is small (the top left corner); on the other hand, our quasars appear to have smaller amplitude than average when \CIV~EQW is small and \CIV~blueshift is large (the bottom right corner). This trend is consistent with the anti-correlation between \sigmadho~and \lledd\ and the suggestions that \CIV~EQW and \CIV~blueshift are \lledd\ indicators \citep{Shemmer2015, rankine2020}. A similar trend was also found by \citet{Rivera2020} using multi-epoch spectroscopy, where a hybrid metric combining \CIV~EQW and \CIV~blueshift was defined to locate quasars along this trend. This hybrid metric was later referred to as the ``\CIV~distance"\revise{~\citep{CIVdistance}} in \citet{Richards2021} and \citet{Rivera2021}---the \CIV~distance increases its value going from the top left corner to the bottom right corner along the distribution of quasars in the \CIV\ parameter space.
Thus, an anti-correlation between \sigmadho~and \CIV~distance should be expected. 

\subsection{The Perturbation Parameters: \tauperturb~and \sigmanoise}

\begin{figure*}[ht!]
    \vspace{.2cm}
    \centering
    \includegraphics[width=0.31\textwidth]{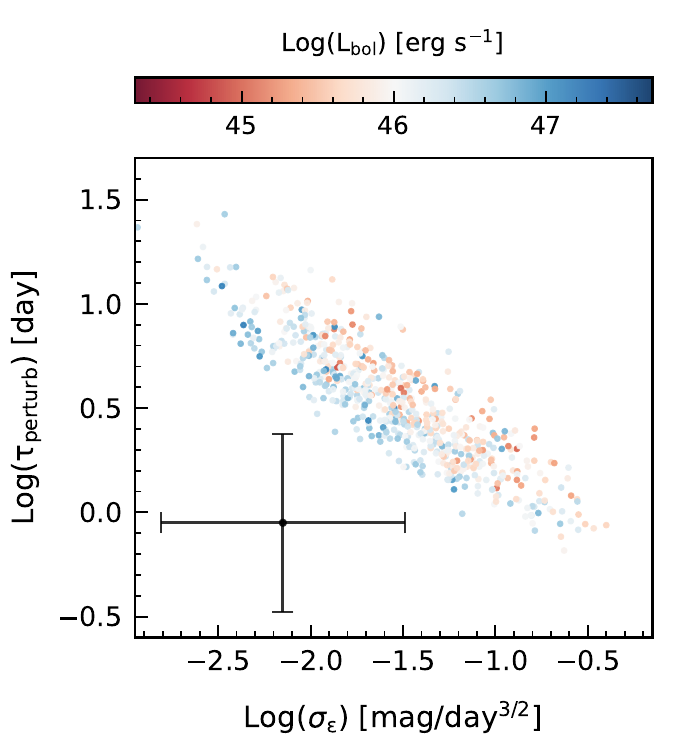}
    \includegraphics[width=0.31\textwidth]{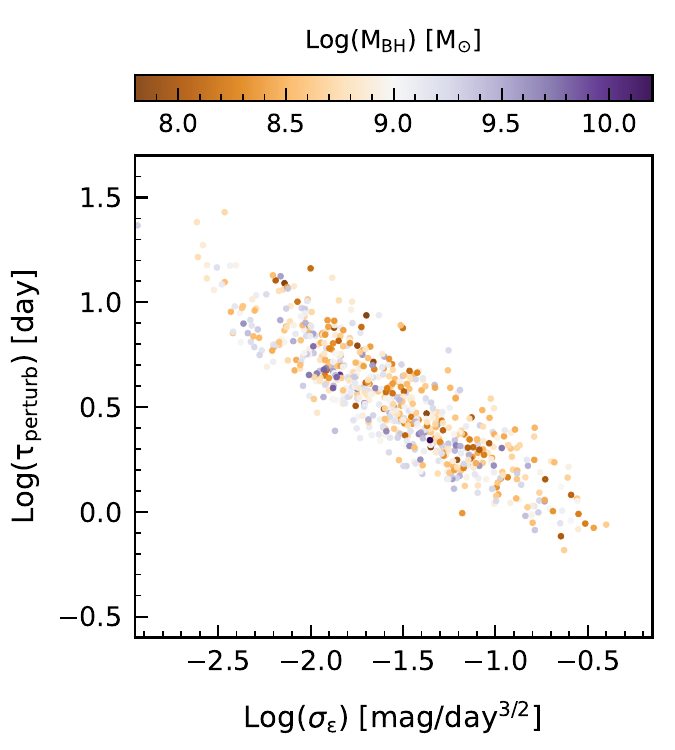}
    \includegraphics[width=0.31\textwidth]{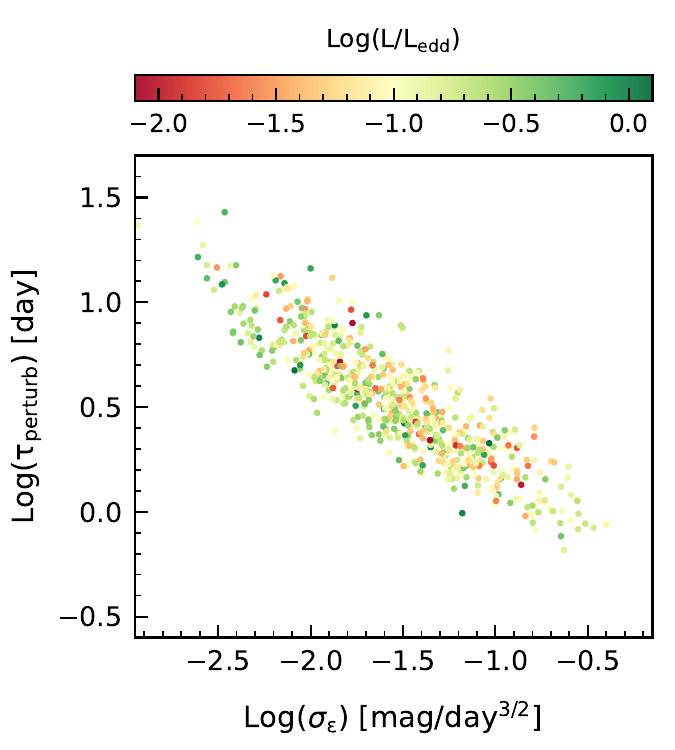}
    \includegraphics[width=0.31\textwidth]{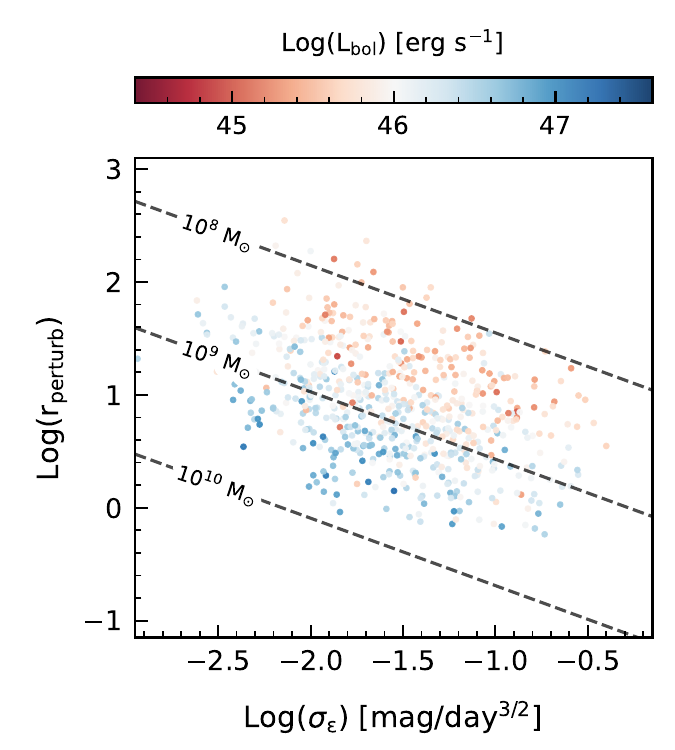}
    \includegraphics[width=0.31\textwidth]{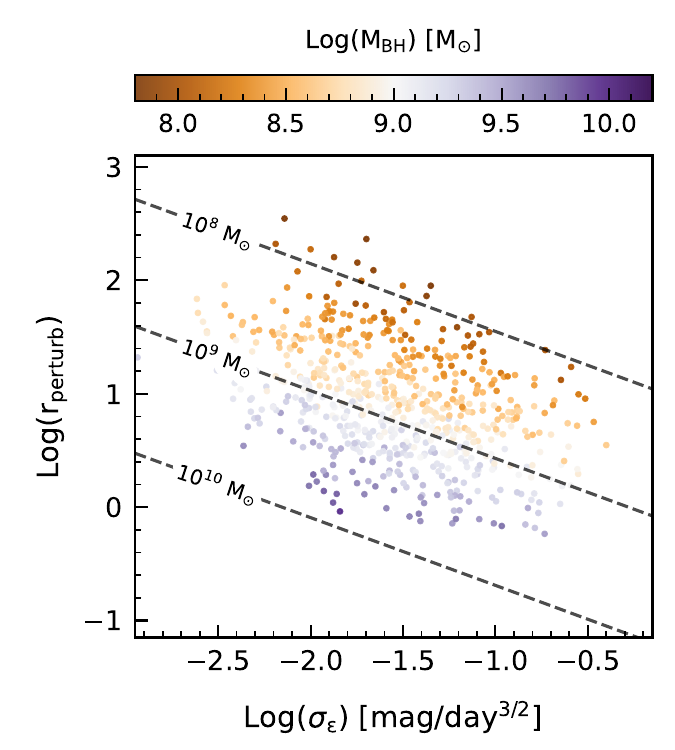}
    \includegraphics[width=0.31\textwidth]{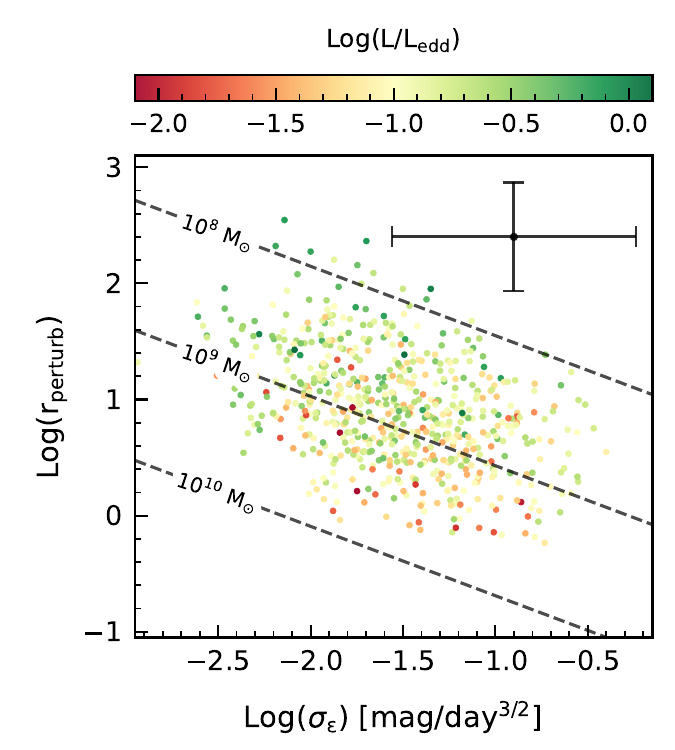}
    \vspace{.2cm}
    \caption{Distribution of quasars from $0.7 < z < 1.9$ with $3.35 <$ log(\lambdarf)~$< 3.45$ in the \tauperturb, \sigmanoise\ parameter space. Both \tauperturb, \sigmanoise\ are corrected to a rest-frame wavelength of 2500 \AA. 
    {\em Top row}: From left to right, quasars are color-coded by their \lbol, \mbh, and \lledd, respectively. The median error bar is displayed at the bottom left corner of the left panel. 
    A trend of decreasing \lbol\ with larger \tauperturb\ and \sigmanoise\ is shown in the left panel. Similar trends for \mbh\ or \lledd\ are not seen in the middle and right panels.
    {\em Bottom row}: Same as the top row but with the y-axis presented in the unit of gravitational radius (\Rg) with $r_{\mathrm{perturb}} = (\tau_{\mathrm{perturb}}*c)/R_{\mathrm{g}}\,\propto\, \tau_{\mathrm{perturb}}/M_{\mathrm{BH}}$. 
    The median error bar is displayed at the top right corner of the right panel.
    The dashed lines are best-fit regression in the form of:~log(\rperturb) $= A*$log(\sigmanoise)~$+~B*$log(\mbh) $+\,C$ for \mbh\ of $10^{8}\,\mathrm{M_{\odot}}$, $10^{9}\,\mathrm{M_{\odot}}$, and $10^{10}\,\mathrm{M_{\odot}}$.
    Weighting \tauperturb\ by \mbh\ reduces the scatter seen in the top row and reveals a new 
    trend of increasing \rperturb\ with increasing \lledd~(see the bottom right panel).
    }    
    \vspace{.0cm}
    \label{fig:perturb_lbol}
\end{figure*}

In the DHO framework, \tauperturb~and \sigmanoise~characterize the driving perturbation to the modeled dynamical system, where \sigmanoise~gives the amplitude of the driving white noise and \tauperturb~describes a characteristic timescale beyond which the perturbation process loses power (see the bottom left panel of Figure~\ref{fig:best_fit_dist} for a reference of the perturbation PSD). In the context of AGN variability modeling, we could expect \tauperturb~and \sigmanoise~to capture the characteristics of the physical mechanisms that drive the observed UV/optical variability, which might correlate with the fundamental properties of AGN.
To that end, we explored the evolution of \lledd, \lbol, and \mbh~across the \tauperturb, \sigmanoise~parameter space. 
Since \tauperturb~and \sigmanoise~do not increase/decrease monotonically with \lambdarf, our analysis only used objects with $3.35 <$ log(\lambdarf)~$< 3.45$, where both \tauperturb~and \sigmanoise~show clear linear dependence on \lambdarf~so that we can calibrate them to the \revise{rest} wavelength of 2500 \AA. As with Section~\ref{subsec:result_amp}, we restrict the sample to $0.7 < z < 1.9$.

The top row of Figure~\ref{fig:perturb_lbol} shows the selected objects ($0.7 < z < 1.9$ and $3.35 <$ log(\lambdarf)~$< 3.45$) in the \tauperturb, \sigmanoise~parameter space, and each panel, from left to right, colors objects by their \lbol, \mbh, and \lledd, respectively.
In the top left panel, we can see a trend of decreasing \lbol~toward the top right corner (\revise{large} \tauperturb\ and \revise{large} \sigmanoise); no apparent evolution of \mbh\ or \lledd\ across the parameter space can be found in the top middle or right panels.

Since the rest-frame \tauperturb~is on the scale of days---comparable to the light-crossing time associated with the size of the accretion disks of our quasars---and since the characteristic radius for emission at fixed wavelengths scale linearly with \mbh\ in the log space~\citep{shakura1973}, it is logical to convert \tauperturb~into a distance scale (\rperturb) expressed in terms of gravitational radius. 
\vspace{-.05cm}
\begin{eqnarray}
    R_{\mathrm{perturb}} &&= \tau_{\mathrm{perturb}}*c, \label{eqn:Rperturb}\\
    R_{\mathrm{g}} &&= GM_{\mathrm{BH}}/c^{2},\label{eqn:Rg}\\
    r_{\mathrm{perturb}} &&= R_{\mathrm{perturb}}/R_{\mathrm{g}}, \label{eqn:rperturb}
\end{eqnarray}
Here, $c$ is the speed of light, $G$ is the gravitational constant, and \mbh~is the mass of the SMBH associated with a quasar. Thus, \rperturb\ is essentially the mass-weighted version of \tauperturb.

The bottom row of Figure~\ref{fig:perturb_lbol} shows the distribution of the selected quasars in the \rperturb, \sigmanoise\ parameter space. 
As with the top three panels of Figure~\ref{fig:perturb_lbol}, from left to right, quasars in each panel are colored by their \lbol, \mbh, and \lledd, respectively. 
As a guide, we plotted the best-fit regression (dashed lines) in the form of: log(\rperturb) $= A*$log(\sigmanoise)$\,+\,B*$log(\mbh)$\,+\,C$ for \mbh\ of $10^{8}\,\mathrm{M_{\odot}}$, $10^{9}\,\mathrm{M_{\odot}}$, and $10^{10}\,\mathrm{M_{\odot}}$.
Note that the scatter seen in the top middle panel is caused by objects with different \mbh\ having the same \tauperturb, weighting \tauperturb\ by \mbh\ (bottom middle panel) enables us to better see the anti-correlation of \tauperturb\ with \sigmanoise\ and facilitate the comparison of different objects on the same physical scale.
A new trend of increasing \rperturb\ with increasing \lledd\ is also revealed in the bottom right panel of Figure~\ref{fig:perturb_lbol}. This new correlation has a non-parametric Spearman rank-order correlation coefficient of \editone{$0.35$} with a two-tailed $p$-value of \editone{$10^{-20}$}.

\begin{figure}[h!]
    \hspace{-1cm}
    \centering
    \plotone{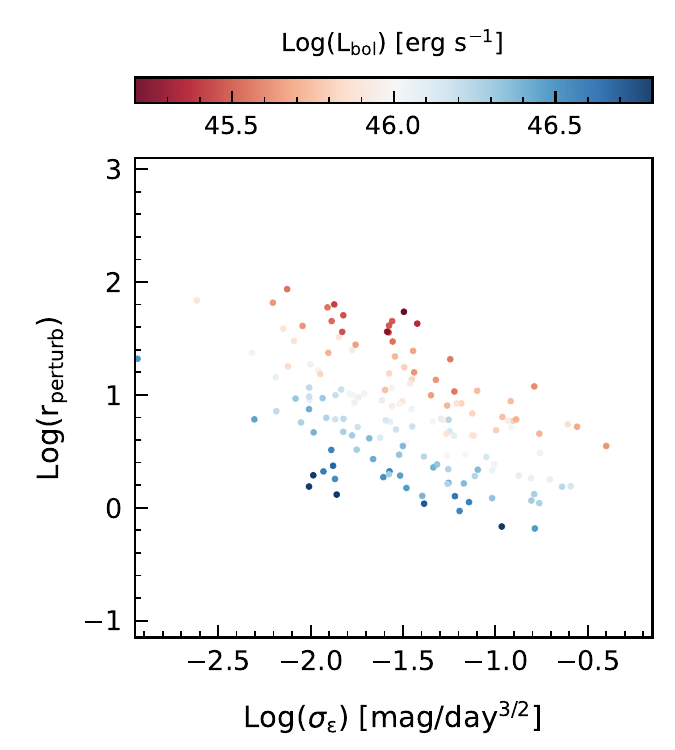}
    \caption{Quasars with $-1.1\,<$ log(\lledd) $<\,-0.9$ in the \rperturb, \sigmanoise\ parameter space. A trend of decreasing \lbol\ towards the top right corner is clearly shown.}
    \label{fig:rperturb_lbol_slice}
\end{figure}

To further elucidate the contributions of \lbol, \mbh, and \lledd\ to the diversity of perturbation parameters shown in Figure~\ref{fig:perturb_lbol}, we selected a subsample of quasars with $-1.1\,<$ log(\lledd) $<\,-0.9$ and plotted their distribution in Figure~\ref{fig:rperturb_lbol_slice} and colored them by their \lbol.
From Figure~\ref{fig:rperturb_lbol_slice} we can see that this subsample spans almost the full range of the distribution shown in the bottom three panels of Figure~\ref{fig:perturb_lbol} and exhibits a much clearer anti-correlation (smoother color gradient) between the perturbation parameters and \lbol~(compared to the bottom left panel in Figure~\ref{fig:perturb_lbol}). The large span of this subsample in the perturbation parameter space and the cleaner/smoother color gradient of \lbol\ suggest that most of the diversity in \rperturb~(and \tauperturb) is driven by \lbol~(or \mbh) and that \lledd\ works independently of \lbol~(or \mbh) in terms of determining the observed distribution and plays only a minor role.

\section{Discussion} \label{sec:discussion}
\subsection{Underdamped DHOs: QPOs and Oscillatory DHOs}

In Section~\ref{sec:results}, we omitted QPOs and oscillatory DHOs from comparing DHO features with the \editone{derived} physical properties of our quasars because of their low classification \revise{precision} (see Table~\ref{tab:dho_class}). However, both subclasses exhibit interesting signatures that can be expected from real physical systems. 

As the name suggests, QPOs vary quasi-periodically (see the simulated light curves in Figure~\ref{fig:best_fit_dist}). Such signals in quasars might be expected from super-massive black hole binaries orbiting each other closely and appearing as a single point source in the image \citep{Begelman1980}. SMBH binaries are expected to emit low-frequency gravitational waves when they merge. QPOs are most likely candidates for those systems and therefore will provide a large pool of potential sources for current and future low-frequency gravitational wave projects \citep{ppta2013, nanograv2013, LISA2017}. \editone{At the same time, persistent quasi-periodic oscillations can also be expected from single SMBHs where the inner accretion flow is geometrically thick and undergoes Lense–Thirring precession~\citep{ingram2009, graham2015}.}

Oscillatory DHOs feature larger \tauperturb\ and smaller \sigmanoise\ compared to overdamped DHOs~(see the top right panel of Figure~\ref{fig:best_fit_dist}). If most of the variability in overdamped DHOs can be attributed to X-ray reprocessing (which will be discussed further in Section~\ref{subsec:discuss_amp}), then the variability revealed in oscillatory DHOs might originate in the local accretion disk. More specifically, the larger \tauperturb~and smaller \sigmanoise\ could be interpreted as characteristics of a perturbation mechanism different than X-ray illumination, e.g., changes in the mass accretion rate, which might features a longer characteristic perturbation timescale (\tauperturb) and a smaller \revise{short-term variability} amplitude (\sigmanoise) \citep{Pereyra2006, li2008, Arevalo2008}. 
Such variability signatures can be expected from quasars with extremely high \lledd~where the radiation from the X-ray corona is weak relative to the intrinsic radiation from the disk \citep{kubota2018} or blocked by a puffed up inner disk (see Figure 15 from \citet{leighly2004} or Figure 18 from \citet{luo2015} for a reference),
both would lead to little or no X-ray reprocessing.

\editone{Besides these extreme objects, we might expect to see a mixture of signatures from both an oscillatory DHO and an overdamped DHO in real light curves, that is, assuming that intrinsic disk variability due to changing mass accretion rate and X-ray reprocessing are contributing comparably to the observed variability~\citep{Arevalo2008}. Such scenarios could also lead to bi-modal distributions in the MCMC samples (see Appendix~\ref{apx:bimodal_mcmc}). In those cases, we will need more advanced modeling tools to decouple the light from different processes. Nonetheless, we have removed objects with bi/multi-modal posterior distributions from our analysis and the analysis presented in Section~\ref{sec:results} is concentrated on the overdamped population only, therefore, it is logical to suspect that the variability features revealed in this investigation are likely dominated by one single mechanism (e.g., X-ray reprocessing).}

\subsection{Long-term Amplitude of Overdamped DHOs: Primarily Determined by \lledd?}\label{subsec:discuss_amp}

We find that the long-term asymptotic variability amplitude of AGN in the overdamped subclass, as characterized by \sigmadho, is anti-correlated with Eddington ratio and black hole mass. This finding is consistent with that resulted from previous investigations utilizing other methods~\citep{vandenberk2004, wilhite2007, macleod2010, ai2010, simm2016}.
The anti-correlation of \sigmadho\ with \lledd\ is expected in a model where the size of the hot X-ray corona relative to the accretion disk anti-correlates with \lledd\ and the optical variability is largely due to reprocessing of X-ray photons~\citep[e.g.,][]{kubota2018, giustini2019}. 
More specifically, a high \lledd~corresponds to a small X-ray corona and a large/strong disk, therefore, less reprocessing of X-ray photons occurs in the disk---leading to a smaller \revise{long-term} variability amplitude; on the other hand, a low \lledd~indicates a large X-ray corona relative to the disk, thus, more reprocessing of X-ray photons occurs in the disk and a larger \revise{long-term} variability amplitude can be expected. 
\editone{We note that the correlations discussed in this section can also be produced by changing mass accretion rate in the disk~\citep{li2008}, moreover, recent work on intensive multi-band AGN reverberation mapping for a handful of objects have provided potential evidence for such scenario~\citep{edelson2017, Starkey2017, edelson2019, cackett2020}, see \citet{cackett2021} for more discussion on this topic}.

If \lledd~sets the basic level of AGN variability, then the additional anti-correlation with \mbh~can be explained by the recognition that the part of the accretion disk that emits at a fixed effective temperature changes radius with increasing/decreasing \mbh. 
The flux \revise{emitted} per unit area at a radius $R >>$ $6$\Rg\ on the disk is defined as
\begin{eqnarray}\label{eqn:disk_bb}
    F(R) = \sigma T(R)^{4}\,&&\propto\,(\dot{m}/M_{\mathrm{BH}})(R/R_{\mathrm{g}})^{-3}\,\nonumber\\
    &&\propto\,\dot{m}M_{\mathrm{BH}}^{2}R^{-3},
\end{eqnarray}
where $T$ is the effective temperature at a radius $R$, $\dot{m}$ is the Eddington ratio, $M_{\mathrm{BH}}$ is the black hole mass, and $R_{\mathrm{g}}$ is the gravitational radius \citep{Novikov1973}. According to this equation, the radius ($R$) on the disk that emits at a fixed temperature (a fixed wavelength), given a known Eddington ratio, moves outwards with increasing \mbh\ resulting in a larger distance to the illuminating X-ray corona. In a `lamppost' geometry, the larger the distance to the X-ray corona, the smaller the intensity of the illuminating X-ray photons and the smaller the variability amplitude due to X-ray reprocessing. 
Therefore, we argue that the variability amplitude could be largely determined by \lledd~with \mbh~acting as a secondary parameter. 


\subsection{Perturbation Parameters of Overdamped DHOs: Indicating a Two-component Accretion Disk?}\label{subsec:warm_disk}

\begin{figure*}[ht!]
    \centering
    \includegraphics[width=.95\textwidth]{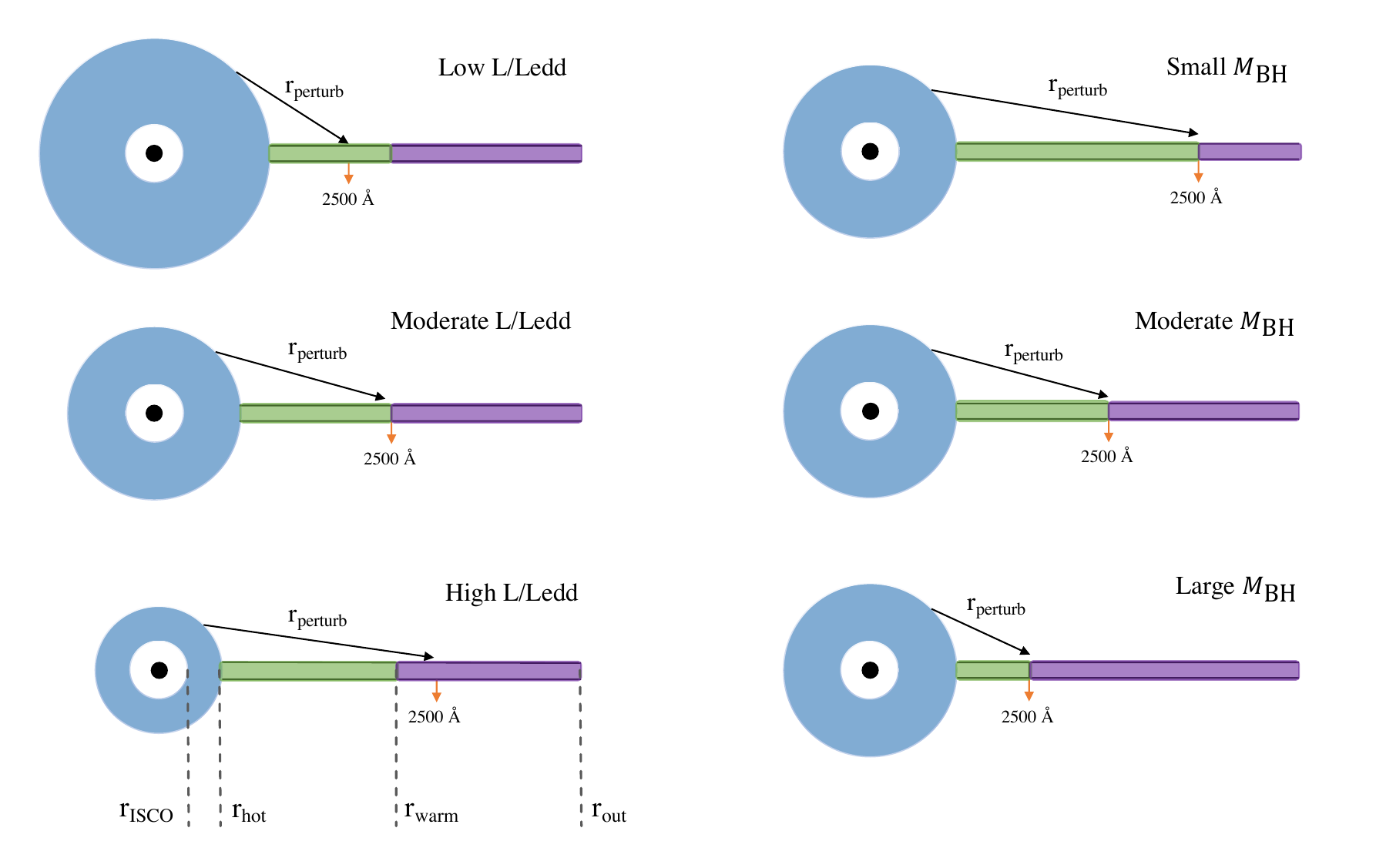}
    \caption{Diagrams providing schematic views of the accretion disk geometry proposed in \citet{kubota2018} for a variety of \lledd\ and \mbh. Note that the diagrams are drawn in the units of gravitational radius \revise{and Equation~\ref{eqn:disk_bb} is used as a reference when discussing the changing disk geometry}. The blue annulus represents the hot X-ray corona, the green slab shows the warm Comptonization region of the disk, and the purple slab corresponds to the standard disk region. The effective emitting wavelength of the disk increases going from $r_{\mathrm{hot}}$ to $r_{\mathrm{out}}$. The \rperturb\ arrow connects the X-ray corona to the disk annulus emitting at an effective wavelength of 2500 \AA.
    {\em Left column}: \mbh\ is held fixed and \lledd\ increases from top to bottom. The size of the X-ray corona decreases and the 2500 \AA\ mark shifts outwards with increasing \lledd. The effective temperature/wavelength of the disk at the boundary between the warm Comptonization region (green slab) and the standard disk region (purple slab) also increases/decreases with \lledd.
    {\em Right column}: \lledd\ is held fixed and \mbh\ increases from top to bottom. When \lledd\ is fixed, the disk temperature (wavelength) at the boundary between the green and purple slab should stay relatively fixed, thus, increasing \mbh\ shifts both $r_{\mathrm{warm}}$ and the 2500 \AA\ disk annulus inwards.
    }
    \label{fig:agn_schem}
\end{figure*}

In Figure~\ref{fig:lambda_dp_all_scatter}, we saw that \tauperturb~and \sigmanoise~do not strictly follow a power-law \revise{relation} with \lambdarf~with a break point at $\approx$2500 \AA; the change in dependency with \lambdarf~could signify that two different physical processes are involved in shaping the observed variability. Our speculation can be explained by a physical model of the accretion flow consisting of: a hot X-ray corona extending from the innermost stable circular orbit (ISCO) to the inner edge of a truncated disk ($r_{\mathrm{hot}} = R_{\mathrm{hot}}/R_{\mathrm{g}}$), a warm Comptonization region (for producing the soft X-ray excess) spanning from $r_{\mathrm{hot}}$ to an intermediate radius ($r_{\mathrm{warm}} = R_{\mathrm{warm}}/R_{\mathrm{g}}$), and a standard cold thin disk going from $r_{\mathrm{warm}}$ to the outer edge of the disk \citep{rozanska2000, czerny2003, sobolewska2004, done2012, kubota2018}. In this model, the warm Comptonization region features two slabs of warm electrons sandwiching the standard disk; it is also assumed that the hot X-ray corona is the main driver of the UV/optical variability. A schematic view of said geometry is shown in Figure~\ref{fig:agn_schem}, where the blue annulus represents the hot X-ray corona, the green slab represents the warm Comptonization region, and the purple slab corresponds to the outer standard disk. Note that the disk does not extend to the ISCO and truncates at $r_{\mathrm{hot}}$.

More concretely, given this two-component accretion disk model (warm Comptonization region + standard thin disk region) and the fact that the effective temperature of a disk annulus scales with its radius (see Equation~\ref{eqn:disk_bb}), we might suspect the observed break point (at $\approx$2500 \AA) in the wavelength dependence of \tauperturb~and \sigmanoise~to correlate with the transition point from the warm Comptonization region to the standard disk region. 
At $r > r_{warm}$ (or \lambdarf~$>$ 2500 \AA\ given our data), the hot X-ray corona directly illuminates the disk, and \tauperturb\ probes the light-crossing time between the emitting disk annulus and the corona.
At $r < r_{warm}$ (or \lambdarf~$<$ 2500 \AA\ given our data), the hot X-ray photons are scattered by the warm electrons in the upper atmosphere of the disk, therefore, the perturbation process exhibits an increased \tauperturb~from that expected for direct illumination (see the best-fit line and the median values in the middle panel of Figure~\ref{fig:lambda_dp_all_scatter}). At the same time, the down-scattered photons (with reduced frequencies) are better absorbed at the disk surface compared to the source X-ray photons because the absorption coefficients for both free-free and bound-free adsorptions are inversely proportional to frequency~\citep{rybicki1986}, which leads to an intensified perturbation amplitude (\sigmanoise). 
We acknowledge that the empirically identified break wavelength of 2500 \AA\ is an ensemble average over the true break wavelengths for our selected quasars and that the break wavelength could shift as a function of \lledd~(see the left column of Figure~\ref{fig:agn_schem}), as suggested by \citet{kubota2018}. However, the size of the error bars in our inferred parameters prohibit us from revealing it. Future investigations utilizing better light curves are needed to characterize that dependency.

The trend revealed by the bottom panel of Figure~\ref{fig:lambda_dp_all_scatter} for \sigmanoise~and the change in behaviors at $\approx$2500 \AA~also agrees with the result presented in \citet{wilhite2005}, which used a composite difference spectrum constructed from $\approx$300 SDSS quasars to demonstrate a similar \lambdarf~dependency of \sigmanoise. 
However, such a trend is not apparent in the correlation of the long-term variability amplitude with \lambdarf---as shown with \sigmadho\ and other metrics~\citep[e.g., \sigmadrw;][]{macleod2010}. 
We suspect the missing imprint of the short-term variability amplitude (\sigmanoise) on the long-term variability amplitude (\sigmadho) could be related to the thermalization of the illuminating photons with the disk over long timescales. 
It is also possible that the correlation of the long-term variability amplitude with \lambdarf\ is itself a function of \lledd\ and/or \lbol\ and that the quasar samples used in previous work and this one for deriving such correlation span a large range of \lledd\ and \lbol. 
Indeed, the quasars used in this work cover a slightly larger range, as characterized by the dispersion of the distribution, of \lledd\ than that used in \citet{wilhite2005}. The \lledd\ distribution of our quasars have a median absolute deviation (MAD) and a IQR of 0.54 dex and 0.27 dex, respectively, whereas the same statistics for quasars used in \citet{wilhite2005} have a value of 0.43 dex and 0.22 dex, respectively. \revise{Moreover, we found \sigmadho\ to follow a similar v-shaped trend with \lambdarf\ as for \sigmanoise~when binned with \lbol\ or \lledd.} A larger sample of quasars with trusted physical properties and high-quality light curves is needed to investigate and characterize how the correlation of \sigmadho\ with \lambdarf\ depends on the value of \lbol\ and \lledd.

Based on the model proposed in \citet{kubota2018} and the fact that we can approximate the effect of the illuminating X-ray corona at $r >> 6$ using a point source on the spin axis at a height of $H = r_{\mathrm{hot}}$~\citep{gardner2017}, \rperturb\ at the break wavelength is effectively $\sqrt{r_{\mathrm{hot}}^2 + r_{\mathrm{warm}}^2}$, assuming it probes the distance from the X-ray corona to the disk at $r_{\mathrm{warm}}$. According to the best-fit parameters for Mrk 509 and PG1115+407 shown in Table 2 of \citet{kubota2018}, $\sqrt{r_{\mathrm{hot}}^2 + r_{\mathrm{warm}}^2}$ should exhibit a negligible correlation with \lledd\ given a fixed \mbh. Indeed, we only see a weak trend of increasing \rperturb\ with increasing \lledd\ from the bottom right panel of Figure~\ref{fig:perturb_lbol}. 
We also suspect that trend as a result of incorrect break wavelength being adopted for quasars having a large range of \lledd, that is, the break wavelength should decrease with increasing \lledd\ and calibrating all \tauperturb\ to the \lambdarf\ of 2500 \AA\ produces a false-positive trend of increasing \rperturb\ with increasing \lledd~(see the diagrams in the left columns of Figure~\ref{fig:agn_schem}). 
Nonetheless, the distance scale reflected by \tauperturb~(\rperturb: 30--150 $R_{\mathrm{g}}$) is comparable to the numbers quoted in Table 2 of \citet{kubota2018}. 
In addition, for a given \lledd, \rperturb\ should decrease with increasing \mbh~(and \lbol), because $r_{\mathrm{warm}}$ (if corresponding to a constant temperature) decreases when \mbh~increases according to Equation~\ref{eqn:disk_bb} (see the right column of Figure~\ref{fig:agn_schem}) and $H = r_{\mathrm{hot}}$ is relatively fixed for a fixed spin~\citep{kubota2018}; the selected subsample of quasars with $-1.1\,<$ log(\lledd) $<\,-0.9$ as shown in Figure~\ref{fig:rperturb_lbol_slice} demonstrates this correlation. 

We currently do not understand the origin of the anti-correlation between \rperturb~and \sigmanoise~(at a fixed \mbh), but it is consistent with a picture where the closer the disk is to the X-ray corona the larger the perturbation amplitude. 
The most straightforward explanation is that the range of spins of the central black holes has produced the diversity in \rperturb/\tauperturb~given all other parameters of the SMBH held fixed; however, further investigation is required to verify this hypothesis. 

Our results lead to conclusions: 1) the warm Comptonization region might be responsible for the observed break in the wavelength dependence of \tauperturb~and \sigmanoise; 2) \tauperturb~at \lambdarf~$>$ 2500 \AA~might be associated with the light crossing time from the X-ray corona to the accretion disk; 3) the short-term variability amplitude decreases with wavelength at \lambdarf~$<$ 2500 \AA~and stays roughy constant (or drops at a much slower rate) with rest\revise{-frame} wavelength---consistent with that shown in \citet{wilhite2005}.

\section{Summary} \label{sec:Summary}

In this work, we have investigated the UV/optical variability of $\approx$12,000 SDSS S82 quasars by modeling their light curves in the $ugriz$ bands as DHO processes. A DHO process can be fully described by four basic parameters: a natural oscillation frequency~($\omega_{0}$), a damping ratio~($\xi$), a characteristic perturbation timescale (\tauperturb), and an amplitude for the perturbing white noise (\sigmanoise). The asymptotic (long-term) variability amplitude is characterized by \sigmadho~(a function of the above four parameters). We explored the correlations of the best-fit DHO parameters and the derived features with the physical properties of our quasars estimated by \citet{shen2011}. The main results are summarized below:

\begin{enumerate}
    \item The distribution of the best-fit DHO parameters splits naturally into two main clusters: an overdamped DHO population and an underdamped DHO population (see Figure~\ref{fig:best_fit_dist}). 
    The overdamped population exhibits similar variability signatures as those characterized by a DRW model. The underdamped population can be further classified (empirically) into the QPO subclass and the oscillatory DHO subclass. Both QPOs and oscillatory DHOs have light curves that are smoother at short timescales (smaller $\xi$) but bumpier at long timescales than those of overdamped DHOs. The QPO subclass features observable quasi-periodicity in its light curve.
    
    
    \item \sigmadho, \tauperturb, and \sigmanoise\ \revise{of overdamped DHOs} evolve with rest-frame wavelength (\lambdarf). \sigmadho~follows a largely monotonic trend with \lambdarf~with a power-law index of \editone{$-0.73~\pm~0.070$}. \tauperturb\ also follows a power-law relation with \lambdarf~but only at \lambdarf\ $>$ 2500 \AA. For \sigmanoise, a \revise{clear} power-law relation is only observed at \lambdarf\ $<$ 2500 \AA. 

    \item After correcting for the wavelength dependence, \sigmadho~(\revise{of overdamped DHOs}) exhibits anti-correlations with both \lledd\ and \mbh---in agreement with the results of previous work~\citep[e.g.,][]{wilhite2007, macleod2010}. However, our best-fit regression (Equation~\ref{eqn:amp_lledd}) suggests steeper anti-correlations of \sigmadho~with \lledd\ and \mbh\ than that reported in \citet{macleod2010} for a DRW model. 

    \item We found that the short-term variability \revise{of overdamped DHOs} as parameterized by \tauperturb\ and \sigmanoise\ is anti-correlated with \lbol.
    
    
    \item We argue that the different characteristics of AGN UV/optical variability revealed \revise{by overdamped DHOs} as listed above
    can be connected together using a physical picture: 1) \lledd~determines the size (extension) of the X-ray corona relative to the \revise{truncated} accretion disk; 2) the majority of the observed variability can be attributed to the reprocessing of X-ray photons by a relatively passive accretion disk; 3) from inside out, the accretion disk is divided into two regions---a warm Comptonization region and a standard thin disk region, where the warm Comptonization region is sandwiched by layers of warm electrons that produce the soft X-ray excess \citep{kubota2018}.
\end{enumerate}

Limited by the cadence and photometric accuracy of the dataset used in this work, we are only able to investigate the correlations of \revise{three features of the overdamped DHO population} with the physical properties of AGN. With light curves of better temporal sampling and/or higher photometric accuracy such as those from current/future time-domain surveys~\citep{chambers2016, ZTF2019a, ivezic2019}, we can further exploit DHO modeling in AGN variability study. For example, the QPO subclass can provide a pool of SMBH binary candidates, when utilized jointly with other SMBH binary discovery methods, both the selection efficiency and completeness can be improved \citep{graham2015, Liu2016, charisi2016}. In addition, oscillatory DHOs are likely candidates for AGN that accrete at extremely high \lledd, where the inner disk has puffed up to prevent the X-ray corona from directly illuminating the disk thus resulting in a smoother light curve at short timescales~\citep{leighly2004, luo2015}. 
Moreover, the discovered anti-correlation of \tauperturb\ and \sigmanoise\ with \lbol, together with the known anti-correlation of \sigmadho\ with \lledd\ and \mbh~(or \lbol), can be utilized to develop new algorithms that will derive \lledd\ for millions of AGN using photometric data alone. Lastly, our sample of quasars span only a limited range in \mbh\ ($10^{8}\,\mathrm{M_{\odot}}$ to $10^{10}\,\mathrm{M_{\odot}}$), applying the same analysis technique used here to AGN of much smaller \mbh\ (e.g., $10^{6}\,\mathrm{M_{\odot}}$) will help further elucidate the correlations between variability and fundamental properties of AGN.

Despite the better sampling and high S/N of light curves coming from current and future time-domain surveys \citep{ZTF2019a, ivezic2019}, we stress the need to develop reliable methods that can effectively merge light curves from multiple surveys for the sake of extending the baseline. 
Given the current likelihood-based inference technique, the long-term variability of AGN (on the scale of years) can only be best constrained when the light curves are much longer than the intrinsic timescales \citep{kozlowski2017a, kozlowski2021}. However, the stochastic nature of AGN variability and its correlation with wavelength make merging light curves by median/mean magnitude a sub-optimal solution. In addition to constructing longer light curves, better inference algorithms that can specifically tackle the effects of sparse sampling and short baseline for light curves will be very helpful. We also emphasize that an ultimate algorithm that can efficiently fit light curves from different bands simultaneously will be essential to perform a `true' CARMA modeling of AGN light curves given that the light curves from different passbands (and the information encoded therein) should be inter-correlated \revise{(see \citet{hu2020} for a heuristic example of such approach)}. 

Last but not least, stochastic diffusion processes like DRW and DHO are statistical models rather than physical models. Therefore, care should be taken when interpreting the timescales extracted from stochastic modeling. One interesting discovery that we made while comparing the DRW features and the DHO features derived from the dataset used in this work is that the decorrelation/decay timescale of DRW ($\tau_{\mathrm{DRW}}$) is not correlated with the decorrelation/decay timescale of DHO (\taudecorr\ or \taudecay) as we would have expected, but instead exhibits a tight correlation with the ratio between \taudecay\ and \tauperturb~\revise{of DHO} (see Figure~\ref{fig:dho_drw}). We discuss plausible origin(s) of this ``mis-match" in Appendix~\ref{apx:dho_drw}.

\acknowledgments
\revise{M.S.V., G.T.R., and J.M. acknowledge support from NASA grant NNX17AF18G.
We thank the referee for a thorough review and the helpful comments. }

Funding for the SDSS and SDSS-II has been provided by the Alfred P. Sloan Foundation, the Participating Institutions, the National Science Foundation, the U.S. Department of Energy, the National Aeronautics and Space Administration, the Japanese Monbukagakusho, the Max Planck Society, and the Higher Education Funding Council for England. The SDSS Web Site is http://www.sdss.org/.

The SDSS is managed by the Astrophysical Research Consortium for the Participating Institutions. The Participating Institutions are the American Museum of Natural History, Astrophysical Institute Potsdam, University of Basel, University of Cambridge, Case Western Reserve University, University of Chicago, Drexel University, Fermilab, the Institute for Advanced Study, the Japan Participation Group, Johns Hopkins University, the Joint Institute for Nuclear Astrophysics, the Kavli Institute for Particle Astrophysics and Cosmology, the Korean Scientist Group, the Chinese Academy of Sciences (LAMOST), Los Alamos National Laboratory, the Max-Planck-Institute for Astronomy (MPIA), the Max-Planck-Institute for Astrophysics (MPA), New Mexico State University, Ohio State University, University of Pittsburgh, University of Portsmouth, Princeton University, the United States Naval Observatory, and the University of Washington.

Funding for the Sloan Digital Sky Survey IV has been provided by the Alfred P. Sloan Foundation, the U.S. Department of Energy Office of Science, and the Participating Institutions. SDSS-IV acknowledges support and resources from the Center for High Performance Computing  at the University of Utah. The SDSS website is www.sdss.org.

SDSS-IV is managed by the Astrophysical Research Consortium for the Participating Institutions of the SDSS Collaboration including the Brazilian Participation Group, the Carnegie Institution for Science, Carnegie Mellon University, Center for Astrophysics | Harvard \& Smithsonian, the Chilean Participation Group, the French Participation Group, Instituto de Astrof\'isica de Canarias, The Johns Hopkins University, Kavli Institute for the Physics and Mathematics of the Universe (IPMU) / University of Tokyo, the Korean Participation Group, Lawrence Berkeley National Laboratory, Leibniz Institut f\"ur Astrophysik Potsdam (AIP),  Max-Planck-Institut f\"ur Astronomie (MPIA Heidelberg), Max-Planck-Institut f\"ur Astrophysik (MPA Garching), Max-Planck-Institut f\"ur Extraterrestrische Physik (MPE), National Astronomical Observatories of China, New Mexico State University, New York University, University of Notre Dame, Observat\'ario Nacional / MCTI, The Ohio State University, Pennsylvania State University, Shanghai Astronomical Observatory, United Kingdom Participation Group, Universidad Nacional Aut\'onoma de M\'exico, University of Arizona, University of Colorado Boulder, University of Oxford, University of Portsmouth, University of Utah, University of Virginia, University of Washington, University of Wisconsin, Vanderbilt University, and Yale University.

This research makes use of the SciServer science platform (www.sciserver.org).
SciServer is a collaborative research environment for large-scale data-driven science. It is being developed at, and administered by, the Institute for Data Intensive Engineering and Science at Johns Hopkins University. SciServer is funded by the National Science Foundation through the Data Infrastructure Building Blocks (DIBBs) program and others, as well as by the Alfred P. Sloan Foundation and the Gordon and Betty Moore Foundation.

\vspace{5mm}
\software{pandas \citep{pandas2010},
         numpy \citep{numpy2020},    
         scipy \citep{scipy2020},
         matplotlib \citep{matplotlib2007},
         astropy \citep{astropy2013, AstropyCollaboration2018},  
         emcee \citep{foreman-mackey2013},
         eztao \citep{Yu2022}
         }

\bibliography{Yu_bib, dho.bib}{}
\bibliographystyle{aasjournal}


\appendix
\section{Bimodal Posterior Distributions of DHO Parameters}

\label{apx:bimodal_mcmc}
\begin{figure}[ht!]
    \centering
    \plottwo{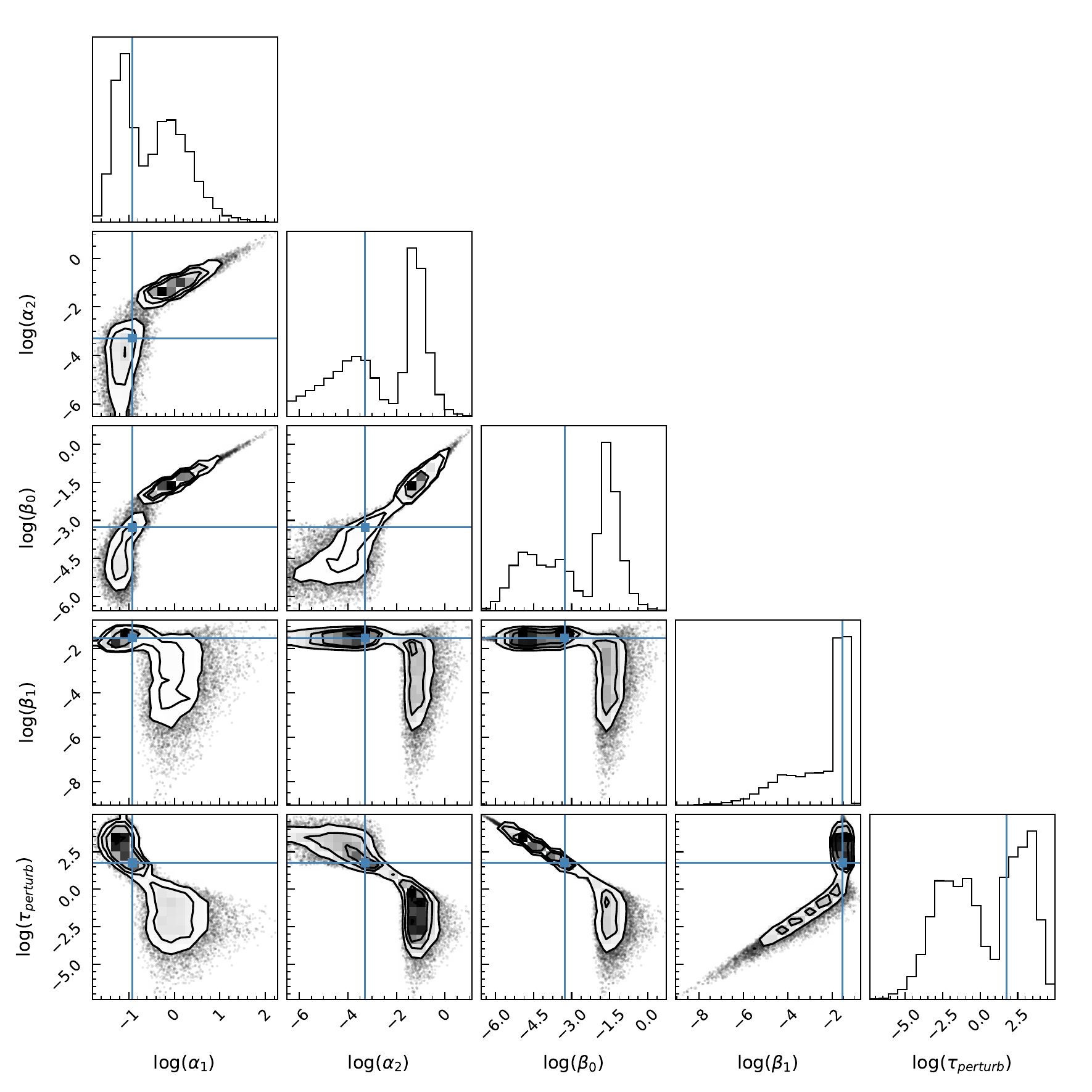}{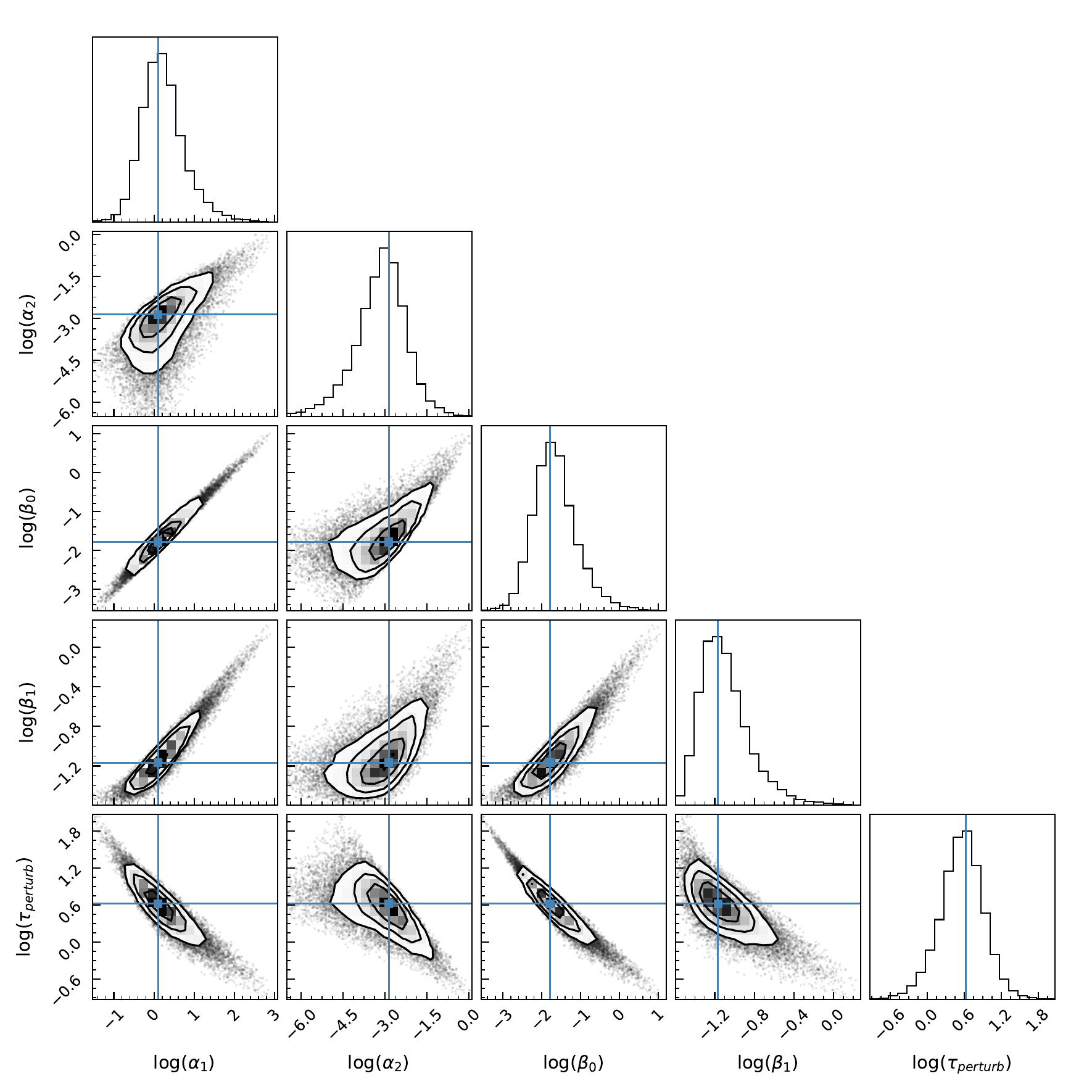}
    \caption{\revise{MCMC posterior distributions for DHO parameters: $\alpha_1$, $\alpha_2$, $\beta_0$, $\beta_1$, and \tauperturb. Shown are DHO fits for $g$-band light curves.
    {\em Left}: A selected example showing a bi-modal distribution; this example has a bi-modality index of 7.25.
    {\em Right}: A selected example showing a posterior distribution with a single mode; this example has a bi-modality index of 1.31.
    }}
    \label{fig:mcmc}
\end{figure}

\begin{figure}
    \centering
    \includegraphics[width=0.6\textwidth]{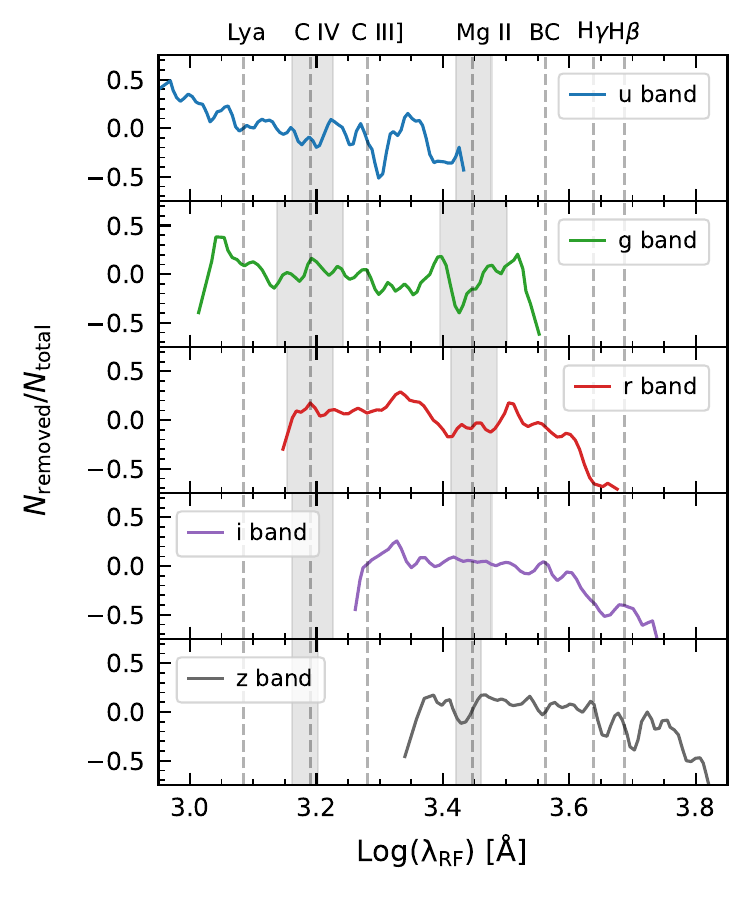}
    \caption{The percentage of DHO fits removed by a cut of \revise{$S_{\mathrm{BM}} > 2.6$} as a function of the rest-frame effective wavelength of each photometric band. The shaded areas show the rest-frame coverage of the photometric bands centered at the \CIV\ and $\mathrm{Mg\,II}$ emission lines. Some local peaks can be spotted at the wavelengths of emission lines (or within the shaded regions), e.g., at the \CIV\ line for $g$ and $r$ bands, and at the $\mathrm{Mg\,II}$ line for $r$ and $z$ bands.
    }
    \label{fig:bm_rate}
\end{figure}

By visually examining the posterior distribution of our MCMC samples, we found that some of them appear to be bi/multi-modal (see the left panel of Figure~\ref{fig:mcmc}). \revise{We suspect this bi/multi-modality to have a mixture of different origins: intrinsic degeneracy (observed variability coming from different physical processes with comparable contributions), the poor sampling of our light curves, and strong emission lines landing in the range of the specific photometric bands.} To automatically remove those fits from our analysis, we designed a bi-modality index $S_{\mathrm{BM}}$, 
\begin{eqnarray}
    &S_{\mathrm{BM}} = \sigma_{\tau_{\mathrm{perturb}}} + \sum\limits_{y \in P}
    [\mathrm{log}(\gamma_{y}) + g(\tau_{y})*\mathrm{log}(\tau_{y}/150)], \quad P = [\alpha_1, \alpha_2, \beta_0, \beta_1]\\
    &\gamma = (p_{75} - p_{25})/(p_{99} - p_{1})\\
    &g(\tau) = \frac{1}{1+e^{-(\tau-150)^{0.2}}}
\end{eqnarray}
\revise{where $\sigma_{\tau_{\mathrm{perturb}}}$ is the ``1-sigma" range of the posterior distribution of DHO's \tauperturb\  in the log scale}, $\tau$ is the auto-correlation time/step for the MCMC (see \citet{foreman-mackey2013}), and $p_n$ is the nth percentile of the MCMC samples. 
Here, $\gamma$ measures the tailness of the marginalized distribution. $S_{\mathrm{BM}}$ penalizes for large $\gamma$ (small tails) and large $\tau$ (MCMC takes more steps to converge), both of which are indicators of potential bi/multi-modal distributions; \revise{at the same time, no incentives are given for small $\tau$ (implemented through $g(\tau)$ as an ``activation function")}. For a Gaussian distribution, $\gamma$ has a nominal value of $\approx$0.3. We also found that the MCMC chains for objects with single-mode posterior distributions (see the right panel of Figure~\ref{fig:mcmc}) usually converge within \revise{150} steps. 
\revise{$\sigma_{\tau_{\mathrm{perturb}}}$ plays the role of identifying unconstrained posteriors. The distribution of $\sigma_{\tau_{\mathrm{perturb}}}$ shows a natural break at 1.4, which is reasonable given that the cadence of our light curves is only sensible on timescales from 1 day to 2000 days---covering approximately 3 dex. That said, DHOs with $\sigma_{\tau_{\mathrm{perturb}}}$ greater than 1.4 (a range of 2.8 dex) should be considered unconstrained.} 
These realizations lead to a selection cut of \revise{$S_{\mathrm{BM}} > 2.6$}, where such DHO fits were not used for the analysis presented in Section~\ref{sec:results}. If the bi/multi-modal distribution is solely caused by emission lines contaminating the continuum variability, we should expect to see a higher percentage of objects being removed by our cut when the rest-frame effective wavelength is around an emission line or the Balmer Continuum (BC), which is also expected to be external to the disk. We investigated this expected \revise{effect} and show the result in Figure~\ref{fig:bm_rate}. From Figure~\ref{fig:bm_rate}, we can see that in each band there are peaks occurring around the wavelengths of selected emission lines (or BC). 
Although not every emission line in every band has a matching local peak \revise{and the peaks are not as prominent}, this is still a supportive evidence for our speculation that emission lines might contribute to the bi/multi-modal posterior distribution shown in Figure~\ref{fig:mcmc}, at least partially. \revise{Emission line variability being a likely source of contamination for our analysis would be a good indicator of the promise of photometric reverberation mapping with the upcoming Rubin C. Observatory Legacy Survey of Space and Time~}\citep{chelouche2012, Chelouche2014, ivezic2019}.

\section{Scaling of \sigmanoise~with redshift}\label{apx:b0_z}

According to Equation~\ref{eqn:dho_amp},  
\begin{equation}
    \sigma_{\mathrm{DHO}}^2 = {\frac{\beta_{1}^{2}\alpha_{2} + \beta_{0}^{2}}{2\alpha_{1}\alpha_{2}}} = \sigma_{\mathrm{\epsilon}}^{2}\,(\frac{\omega_{0}^{2}\tau_{\mathrm{perturb}}^{2} + 1}{2\xi\omega_{0}^{3}}).
\end{equation}
where $\omega_0 \propto T^{-1}$, \tauperturb~$\propto T$ ($T$ is a timescale), and $\xi$ is unitless. Since variability amplitude (\sigmadho) does not scale with redshift ($z$), we have,
\begin{equation}
    \beta_0^2 = \sigma_{\mathrm{\epsilon}}^{2} = \sigma_{\mathrm{DHO}}^2 {(\frac{2\xi\omega_{0}^{3}}{\tau_{\mathrm{perturb}}^{2}\omega_{0}^{2} + 1})} \ \propto\ T^{-3}\ \propto\ (1 + z)^{3}
\end{equation}
which gives us $\beta_0 \propto (1 + z)^{3/2}$.

\section{Overdamped DHO vs. DRW}\label{apx:dho_drw}

\begin{figure}[ht!]
    \centering
    \includegraphics[width=0.46\textwidth]{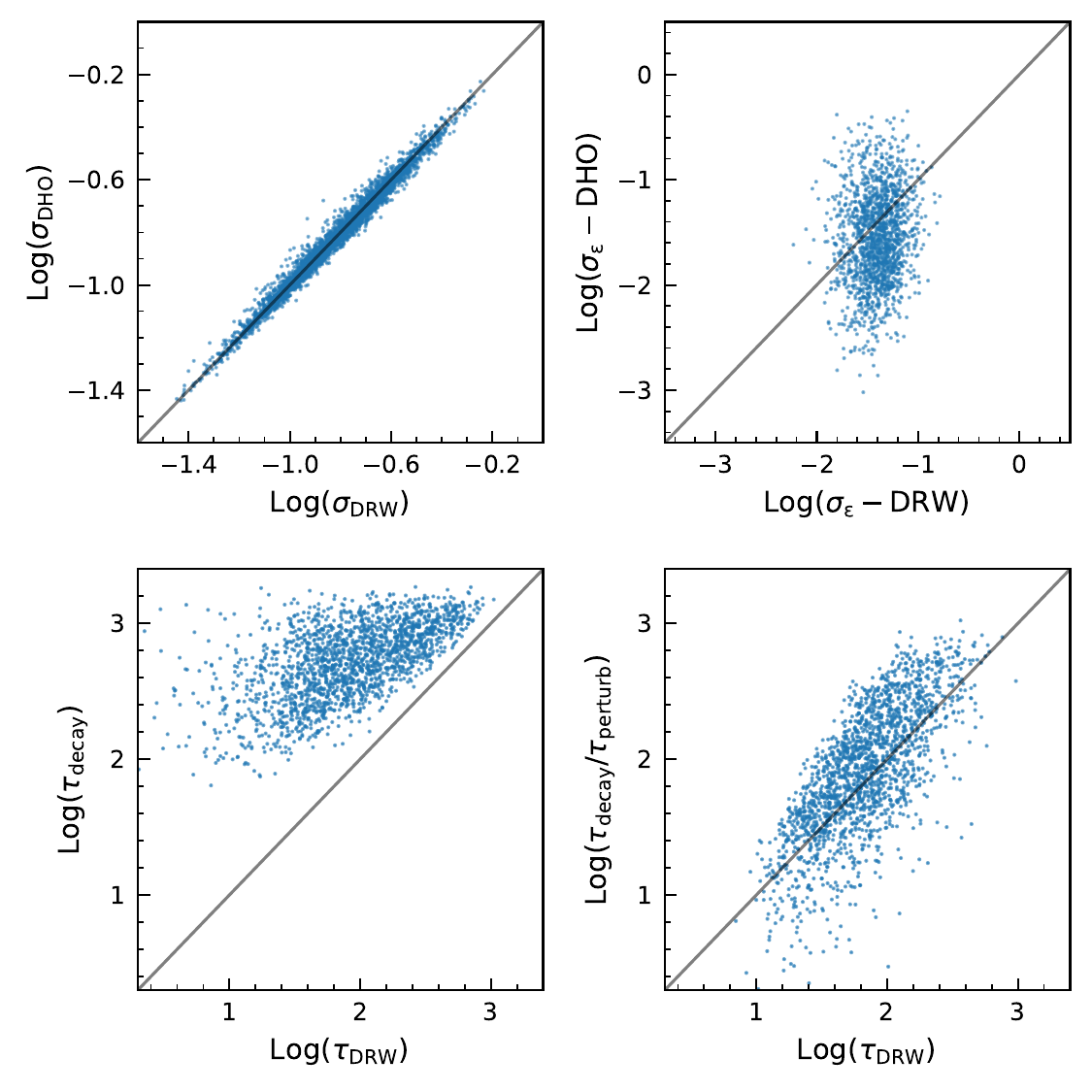}
    \caption{A comparison of similar variability signatures extracted by DHO and DRW. The black solid lines in each panel show the one-to-one correspondence.
    {\em Top left}: \sigmadho\ vs. \sigmadrw, both characterize the asymptotic variability amplitude of the modeled system.
    {\em Top right}: Driving white noise amplitude (\sigmanoise) of DHO and of DRW. 
    {\em Bottom left}: \taudecay~of DHO vs.~$\tau_{\mathrm{DRW}}$, both report the e-folding decay timescale of its auto-correlation function.
    {\em Bottom right}: \taudecay/\tauperturb~of DHO vs.~$\tau_{\mathrm{DRW}}$. 
    }
    \label{fig:dho_drw}
\end{figure}

The \revise{overdamped} DHO model is similar to a DRW, but is more flexible. It would be interesting to compare the variability information extracted by the two models side by side using a same dataset. DRW can be generally characterized by an asymptotic amplitude (\sigmadrw) and a long-term decay timescale ($\tau_{\mathrm{DRW}}$). The features extracted by DHO that are comparable to those two DRW parameters are \sigmadho~and \taudecay. We fitted both models to the $g$-band light curves of our quasars. 
We compared \sigmadho~with \sigmadrw, \sigmanoise\ (the $\beta_0$ coefficient in Equation~\ref{eqn:dho1} and Equation~\ref{eqn:drw}) of DHO with that of DRW, and \taudecay\ of DHO with $\tau_{\mathrm{DRW}}$; the results are shown in the top left, top right, and bottom left panel of Figure~\ref{fig:dho_drw}, respectively. 
From those three comparisons, we only found a good one-to-one correlation between \sigmadho~and \sigmadrw. One would have expected \taudecay\ of DHO to correlate with $\tau_{\mathrm{DRW}}$, since both characterize how long it takes for the modeled system to forget about its past self, however, we did not see that. Instead, we found a strong correlation between the ratio of \taudecay\ to \tauperturb~and $\tau_{\mathrm{DRW}}$ (see the bottom right panel of Figure~\ref{fig:dho_drw}). The slight ``overestimation" of \taudecay/\tauperturb~at large $\tau_{\mathrm{DRW}}$ can be explained by the well known underestimation of $\tau_{\mathrm{DRW}}$ when the light curve span is shorter than 10 times the intrinsic $\tau_{\mathrm{DRW}}$ \citep{kozlowski2017a}, which is exactly what is being shown. 
We are not certain about the origin of this correlation, but if \tauperturb~is physical then we suspect that the DRW model might be overlooking this information. More specifically, since DRW requires the perturbation process to be a white-noise with a flat power spectrum, then $\tau_{\mathrm{DRW}}$ might be in units of \tauperturb~because the DHO perturbation process only behaves like a white-noise at a timescale longer than \tauperturb~(see the bottom left panel of Figure~\ref{fig:best_fit_dist} for the power spectrum density of the DHO perturbation process). 
\revise{In addition, studies utilizing the power spectrum density technique have also suggested a potential second short-term characteristic timescale (relative to the known long-term decorrelation timescale) in AGN light curves, where the reported short-term timescales are comparable to the \tauperturb\ revealed in this work}~\citep{zu2013, Stone2022}.
We stress that the correlation shown between \taudecay/\tauperturb~and $\tau_{\mathrm{DRW}}$ does not necessarily invalidate a DRW (or DHO) description of AGN variability, but argues for more careful examinations of the timescales returned by such modeling. 

\end{document}